\documentclass[traditabstract,bibyear]{aa}
\usepackage{epsfig}
\usepackage{amsmath}
\usepackage{subfigure}
\usepackage{natbib}
\usepackage{multirow}
\usepackage{color}
\usepackage{longtable}
\usepackage{graphicx}
\usepackage{epstopdf}
\usepackage{booktabs}
\usepackage{txfonts}
\newcommand{\jms}{J.~Mol.~Spectrosc.}   
  
\newcommand{\jmst}{J.~Mol.~Struct.}

\newcommand{\kms}{km s$^{-1}$}

\begin{document}

\title{
CN and CCH derivatives of ethylene and ethane: Confirmation of the detection of CH$_3$CH$_2$CCH in TMC-1 
\thanks{Based
on observations with the Yebes 40m radio telescope (projects 19A003, 20A014, 20D023, 21A011, 21D005, 
22A007, 22B029, and 23A024) and the IRAM 30m radio telescope. The 40m radio telescope at Yebes Observatory is operated by the Spanish Geographic Institute (IGN, Ministerio de Transportes y Movilidad Sostenible).
IRAM is supported by INSU/CNRS (France), MPG (Germany), and IGN (Spain).}}

\author{
J.~Cernicharo\inst{1},
B.~Tercero\inst{2,3},
M.~Ag\'undez\inst{1},
C.~Cabezas\inst{1},
R.~Fuentetaja\inst{1},
N.~Marcelino\inst{2,3}, and
P.~de~Vicente\inst{3}
}

\institute{Dept. de Astrof\'isica Molecular, Instituto de F\'isica Fundamental (IFF-CSIC),
C/ Serrano 121, 28006 Madrid, Spain. \newline \email jose.cernicharo@csic.es \& marcelino.agundez@csic.es
\and Observatorio Astron\'omico Nacional (OAN, IGN), C/ Alfonso XII, 3, 28014, Madrid, Spain.
\and Centro de Desarrollos Tecnol\'ogicos, Observatorio de Yebes (IGN), 19141 Yebes, Guadalajara, Spain.
}

\date{Received: 07/02/2024; Accepted: 08/03/2024}

\abstract{We present a study of CH$_3$CH$_2$CCH, CH$_3$CH$_2$CN, CH$_2$CHCCH, and CH$_2$CHCN in TMC-1
using the QUIJOTE$^1$ line survey. We confirm the presence of CH$_3$CH$_2$CCH in TMC-1, which was previously reported 
as tentative by our group. From a detailed study of the ethynyl and cyanide derivatives of CH$_2$CH$_2$ and CH$_3$CH$_3$ in TMC-1,
we found that the CH$_2$CHCCH/CH$_2$CHCN and CH$_3$CH$_2$CCH/CH$_3$CH$_2$CN abundance ratios are 1.5$\pm$0.1 and 4.8$\pm$0.5,
respectively. 
The derived CH$_2$CHCCH/CH$_3$CH$_2$CCH abundance ratio is 15.3$\pm$0.8,
and that of CH$_2$CHCN over
CH$_3$CH$_2$CN is 48$\pm$5. All the single substituted isotopologs of vinyl cyanide have been detected, and we found that the first and second carbon substitutions in CH$_2$CHCN
provide a $^{12}$C/$^{13}$C ratio in line with
that found for other three-carbon bearing species such as HCCNC and HNCCC. However, the third $^{13}$C isotopolog, CH$_2$CH$^{13}$CN, presents an increase
in its abundance similar to that found for HCCCN. Finally, we observed eight $b$-type transitions of CH$_2$CHCN,
and we find that their intensity cannot be fitted adopting the dipole moment $\mu_b$ derived previously. These transitions
involve the same rotational levels as those of the $a$-type transitions. From their intensity, we obtain
$\mu_b$=0.80$\pm$0.03\,D, which is found to be in between earlier values derived
in the laboratory using
intensity measurements or the Stark effect. Our chemical model indicates that the abundances of 
CH$_3$CH$_2$CCH, CH$_3$CH$_2$CN, CH$_2$CHCCH, and CH$_2$CHCN observed in TMC-1 can be explained in terms 
of gas-phase reactions.}


\titlerunning{CH$_3$CH$_2$CH and CH$_3$CH$_2$CN in TMC-1}
\authorrunning{Cernicharo et al.}

\maketitle

\section{Introduction}

The ultra-sensitive line survey 
QUIJOTE\footnote{\textbf{Q}-band \textbf{U}ltrasensitive \textbf{I}nspection \textbf{J}ourney
to the \textbf{O}bscure \textbf{T}MC-1 \textbf{E}nvironment}
performed with the Yebes 40m radio telescope
towards the starless cold core TMC-1 \citep{Cernicharo2021a}, together with an ultra-deep line survey of the carbon-rich star IRC+10216 \citep{Pardo2022}, has recently permitted the unambiguous detection of nearly 70 new molecules in space \citep[][and references therein]{Cernicharo2021a,Pardo2021,Cernicharo2023a,Cabezas2023,Cernicharo2023b}. Together with
the discoveries of the GOTHAM line survey on TMC-1 \citep{McGuire2018}, 
and the IRAM 30m and Yebes 40m survey of G+0693-0.027 toward the Galactic center \citep[see, e.g.,][]{Rivilla2023}, 
the
number of molecules discovered in space has increased from 200 to around 300 species
in the past four years, $\sim$70 of which lie in TMC-1. These results place 
strong constraints on the chemical networks and models of interstellar clouds and on our view on the formation of aromatic rings in these cold objects \citep{McGuire2018,McGuire2021,Cernicharo2021b,Cernicharo2023c,Agundez2023a}. 

TMC-1 is known to harbor abundant cyanopolyynes and the unsaturated
carbon-chain radicals C$_n$H and C$_n$N. 
Even vibrationally excited C$_6$H has been found
in TMC-1 with the QUIJOTE line survey \citep{Cernicharo2023d}.
Nevertheless, the chemistry of this cold core also produces large abundances for more saturated hydrocarbon 
species such as CH$_3$CHCH$_2$ \citep{Marcelino2007}, 
CH$_3$CCH \citep[see, e.g.,][and references threin]{Cabezas2021,Agundez2021},
CH$_2$CCH \citep{Agundez2021,Agundez2022}, vinyl acetylene \citep{Cernicharo2021b}, benzyne \citep{Cernicharo2021a}, 
cyclopentadiene \citep{Cernicharo2021c}, and indene \citep{Cernicharo2021c,Burkhardt2021}. 
In addition, the presence of benzene and naphthalene has been inferred from the detection of their CN 
and CCH derivatives \citep{McGuire2018,McGuire2021,Loru2023}. Recently, the CN
functionalized forms of indene were also found in the same source by \citet{Sita2022}.

In this Letter, we present a systematic study of the ethynyl and cyanide derivatives 
of ethylene (CH$_2$CH$_2$) and ethane (CH$_3$CH$_3$) in TMC-1, confirming the previous tentative
identification of CH$_3$CH$_2$CCH \citep{Cernicharo2021b}. We also report abundances for the singly $^{13}$C, $^{15}$N, and D substituted isotopologs of CH$_2$CHCN.

\section{Observations}

The observational data used in this work are part of QUIJOTE \citep{Cernicharo2021a}, 
which is a spectral line survey of TMC-1 in the Q band carried out with the Yebes 40m telescope at 
the position $\alpha_{J2000}=4^{\rm h} 41^{\rm  m} 41.9^{\rm s}$ and $\delta_{J2000}=
+25^\circ 41' 27.0''$, corresponding to the cyanopolyyne peak (CP) in TMC-1. The receiver 
was built within the Nanocosmos project\footnote{\texttt{https://nanocosmos.iff.csic.es/}} 
and consists of two cold high-electron mobility transistor amplifiers that cover the 
31.0-50.3 GHz band with horizontal and vertical polarizations. The receiver temperatures 
achieved in the 2019 and 2020 runs vary from 22 K at 32 GHz to 42 K at 50 GHz. Some 
power adaptation in the down-conversion chains have reduced the receiver temperatures 
during 2021 to 16\,K at 32 GHz and 30\,K at 50 GHz. The backends are 
$2\times8\times2.5$ GHz fast Fourier transform 
spectrometers with a spectral resolution of 38 kHz, providing the whole coverage 
of the Q band in both polarizations.  A more detailed description of the system 
is given by \citet{Tercero2021}. 

The data of the QUIJOTE line survey presented here were gathered in several 
observing runs between November 2019 and July 2023.  
All observations were performed using the frequency-switching observing mode with 
a frequency throw of 8 and 10 MHz. The total observing time on the source 
for data taken with frequency throws of 8 MHz and 10 MHz was 737 and 465 hours, 
respectively. Hence, the total observing time on source was 1202 hours. The 
measured sensitivity varied between 0.07 mK (70 $\mu$K) at 32 GHz and 0.2 mK at 49.5 GHz.  
The sensitivity of QUIJOTE is about 50 times better than that of previous 
line surveys in the Q band of TMC-1 \citep{Kaifu2004}. For each frequency 
throw, different local oscillator frequencies were used in order to remove 
possible side-band effects in the down-conversion chain. A detailed description 
of the QUIJOTE line survey is provided in \citet{Cernicharo2021a}.
The data analysis procedure has been described by \citet{Cernicharo2022}.

The averaged main-beam efficiency measured during our observations  
varied from 0.66 at 32.4 GHz to 0.50 at 48.4 GHz \citep{Tercero2021} and can be given across the Q band by
$B_{\rm eff}$=0.797 exp[$-$($\nu$(GHz)/71.1)$^2$]. The averaged
forward telescope efficiency is 0.97.
The telescope beam size at half-power intensity is 54.4$''$ at 32.4 GHz and 36.4$''$ 
at 48.4 GHz. 

We also included in this work data from the 3\,mm line survey performed with the IRAM 30m telescope. These data cover the 
full available band at the telescope, between 71.6 GHz and 
117.6 GHz. The EMIR E0 receiver was connected to the Fourier Transform Spectrometers 
(FTS) in its narrow mode. The FTS provide a spectral resolution of 49 kHz and a total 
bandwidth of 7.2 GHz per spectral setup. Observations were performed in several runs. 
Between January and May 2012, we completed the scan 82.5--117.6 GHz \citep{Cernicharo2012b}. 
In August 2018, after the upgrade of the E090 receiver, we extended the survey down 
to 71.6 GHz. More recent high-sensitivity observations in 2021 were used to improve 
the signal-to-noise ratio (S/N) in several frequency windows \citep{Agundez2022,Cabezas2022}. 
The final 3mm line survey has a sensitivity of 2-10 mK. However, at some selected 
frequencies, the sensitivity is as low as 0.6 mK. All the observations were performed using the 
frequency-switching method with a frequency throw of 7.14 MHz. The IRAM 30m beam 
varied between 34$''$ and 21$''$ at 72 GHz and 117 GHz, respectively, while the beam 
efficiency took values of 0.83 and 0.78 at the same frequencies, following the 
relation B$_{eff}$= 0.871 exp[$-$($\nu$(GHz)/359)$^2$]. The forward efficiency at 3\,mm is 0.95.

The intensity scale 
used in this study is the antenna temperature ($T_A^*$). 
Calibration was performed using two absorbers at 
different temperatures and the atmospheric transmission model ATM \citep{Cernicharo1985, 
Pardo2001}. The absolute calibration uncertainty was 10\,$\%$. However, the relative
calibration between lines within the QUIJOTE survey is certainly better as all lines
were observed simultaneously.
The data were analyzed with the GILDAS package\footnote{\texttt{http://www.iram.fr/IRAMFR/GILDAS}}.

\begin{figure}
\centering
\includegraphics[scale=0.455]{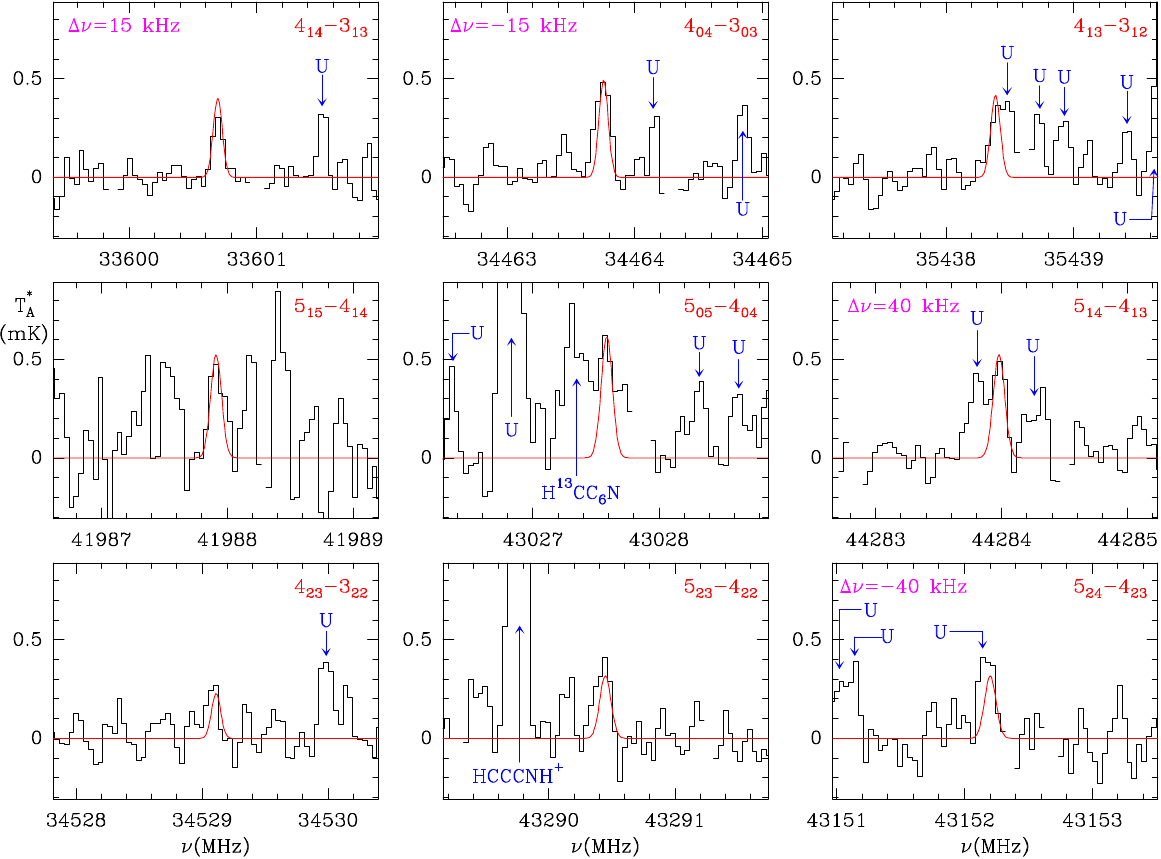}
\caption{Observed lines of CH$_3$CH$_2$CCH in TMC-1. 
The line parameters are given in Table \ref{line_parameters}.
The abscissa corresponds to the rest frequency assuming a velocity for the source of 5.83 km\,s$^{-1}$. 
The ordinate is the antenna temperature corrected for atmospheric and telescope losses in mK. 
Quantum numbers are indicated in the top right corner of each panel.
The red line corresponds to the synthetic spectrum derived from the LTE model described in Sect. \ref{sec:CH3CH2CCH}. 
Blanked channels correspond to negative features produced in the folding of the frequency-switching data.
For some lines, we found a frequency shift of up to $\pm$40 kHz with respect to the predictions (see Sect. \ref{sec:CH3CH2CCH}). This 
is indicated in
magenta in the corresponding panels.
}
\label{fig:CH3CH2CCH}
\end{figure}

\section{Results}\label{sec:results}

The line identification was performed using the MADEX code \citep{Cernicharo2012} and the 
CDMS and JPL catalogs \citep{Muller2005,Pickett1998}.
The intensity scale 
used in this study is the antenna temperature ($T_A^*$). Consequently, the telescope parameters and source 
properties were
used to model the emission of the different species to produce synthetic spectra in this
temperature scale. The source was assumed to be circular
with a uniform brightness temperature and a radius of 40$''$ \citep{Fosse2001}.
The procedure to derive line parameters is described in Appendix \ref{app:lineparameters}.
To model the observed line intensities, MADEX uses a local thermodynamical equilibrium (LTE) 
hypothesis supported by rotational diagrams or adopts a large
velocity gradient approach (LVG). In the later case, MADEX uses the formalism described by \citet{Goldreich1974}.
Unfortunately, no collisional rates are available for the species studied in this work.
The permanent dipolar moments and spectroscopic sources for the molecular species observed in this work 
are discussed in the next sections.

\subsection{CH$_3$CH$_2$CCH (ethyl acetylene)}\label{sec:CH3CH2CCH}

In spite of the large observed abundance of CH$_3$CH$_2$CN in hot cores and hot corinos (see Sect. \ref{sec:CH3CH2CN}), 
the homologous acetylenic form, CH$_3$CH$_2$CCH, has not been reported in this type of environment so far. The detection is less favorable 
due to the modest dipole moment \citep[$\mu_a$=0.763\,D and $\mu_b$=0.17\,D;][]{Landsberg1983}
compared to that of the cyanide form (see Sect. \ref{sec:CH3CH2CN}).
We recently reported a tentative detection of this species toward the cold starless core TMC-1. The claim was based on 
the observation of a few lines and a spectral stacking of all
the transitions in the QUIJOTE line survey \citep{Cernicharo2021b}. Here, we present the detection of nine lines
of the molecule, including some $K_a$=2 transitions. The data are shown in Fig. \ref{fig:CH3CH2CCH}, and their line
parameters are given in Table \ref{line_parameters}. The laboratory data for the rotational spectroscopy of this
species were summarized by \citet{Steber2012}. We note, however, that in the microwave domain, \citet{Bestmann1985} reported a small
internal rotation splitting of about 100 kHz, but the uncertainty of their measurements
was on the same order. Although the laboratory data cover high quantum numbers (up to $J$=46 and $K$=28) and 
the frequency predictions should be accurate enough in the QUIJOTE band, our
observations show  frequency shifts with respect to the predictions of up to 40 kHz (our channel width) for four transitions (indicated in magenta in Fig. \ref{fig:CH3CH2CCH} in the
top left corner of the panels).
This difference is compatible with the uncertainties of the laboratory measurements in the Q band 
and with the possible splitting between the $A$ and $E$ symmetry species at
these frequencies. To derive
the column density of this molecule, we assumed a rotational temperature of 9\,K. This assumption was based on the low dipole moment of the
molecule and on the results we obtained for CH$_2$CHCCH 
\citep[see Sect. \ref{sec:CH2CHCCH} and][]{Cernicharo2021b}. The derived value is
$N$\,=\,(6.2$\pm$0.2)$\times$10$^{11}$ cm$^{-2}$ (see Table \ref{columndensities}). Using the
column density derived in Sect. \ref{sec:CH2CHCCH} for vinyl acetylene, we obtain a
CH$_2$CHCCH/CH$_3$CH$_2$CCH abundance ratio of 15.3$\pm$0.8.

\begin{figure}
\centering
\includegraphics[scale=0.455]{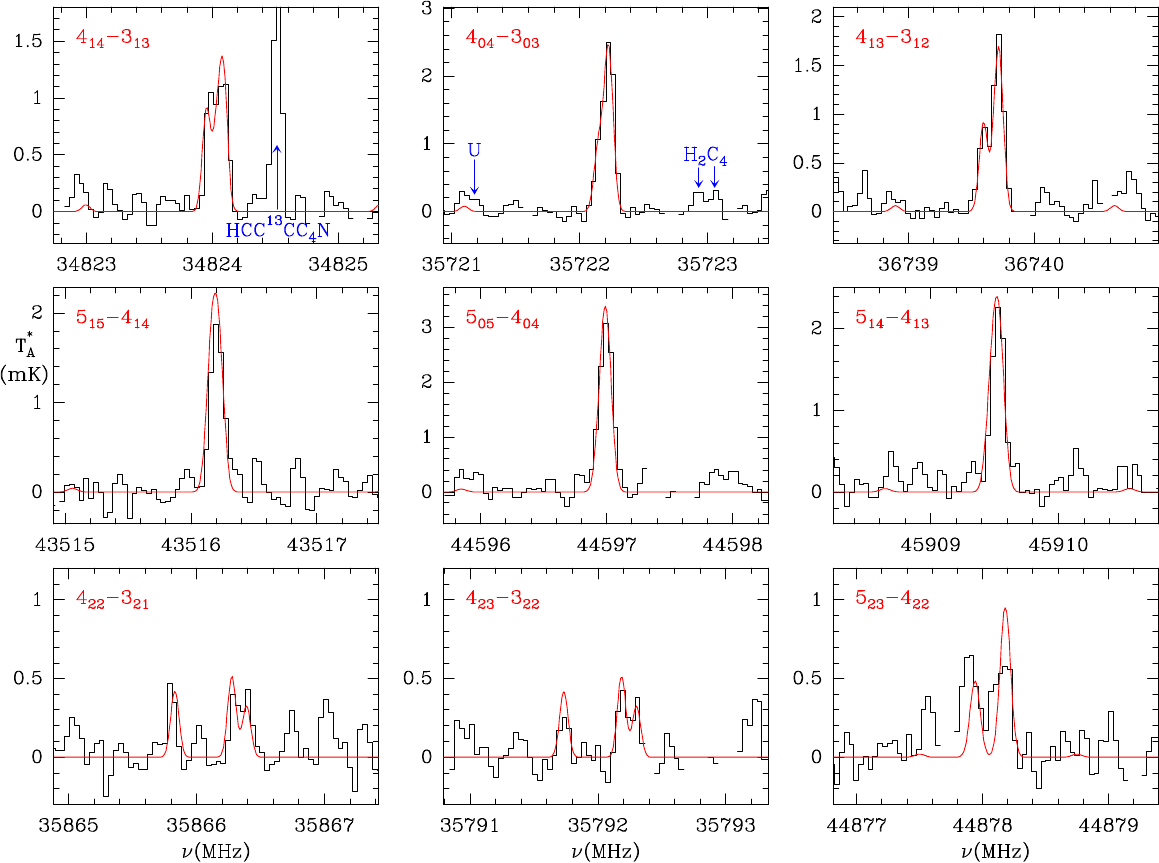}
\caption{Observed lines of CH$_3$CH$_2$CN in TMC-1. 
The line parameters are given in Table \ref{line_parameters}.
The abscissa corresponds to the rest frequency assuming a velocity for the source of 5.83 km\,s$^{-1}$. 
The ordinate is the antenna temperature corrected for atmospheric and telescope losses in mK. 
Quantum numbers are indicated in the top left corner of each panel.
The red line corresponds to the synthetic spectrum derived from the model described in Sect. \ref{sec:CH3CH2CN}. 
Blanked channels correspond to negative features produced in the folding of the frequency-switching data.
}
\label{fig:CH3CH2CN}
\end{figure}

\subsection{CH$_3$CH$_2$CN (ethyl cyanide)}\label{sec:CH3CH2CN}

\begin{figure}
\centering
\includegraphics[scale=0.6]{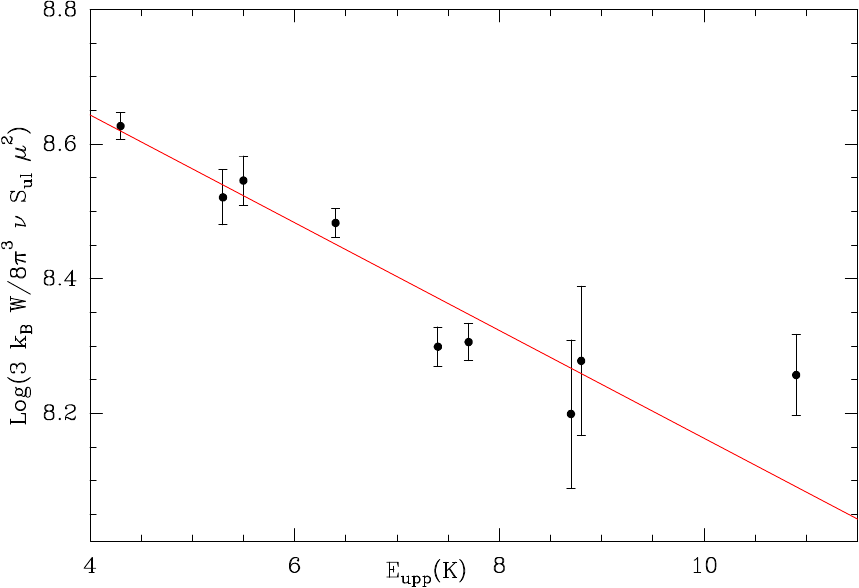}
\caption{Rotational diagram of the observed lines of CH$_3$CH$_2$CN from the data of Table \ref{line_parameters}.
A rotational temperature of 5.5$\pm$0.5\,K is derived.}
\label{fig:CH3CH2CN_Trot}
\end{figure}

CH$_3$CH$_2$CN and several of its vibrationally excited states exhibit a dense spectrum in warm molecular clouds such as Orion-KL and SgrB2 
\citep{Johnson1977,Gib2000,Daly2013,Margules2018}. However, the molecule was not detected
in cold dark clouds until recently, when we reported several of its rotational lines 
toward TMC-1 \citep{Cernicharo2021b}. In this work, we present the spectra for this species using
the latest QUIJOTE data. The spectroscopic laboratory frequency measurements are from \citet{Fukuyama1996} 
and \citet{Brauer2009},
from which we
derived the rotational and distortion constants and implemented them in MADEX.
Nine $a$-type lines were detected, and they are shown in Fig. \ref{fig:CH3CH2CN}, while their
line parameters are given in Table \ref{line_parameters}. Some of the $K_a$=1 and all $K_a$=2 lines show hyperfine structure. 
The frequencies including hyperfine structure were adopted from the CDMS catalog \citep{Muller2005}.
A rotational 
diagram analysis of the data provided a rotational temperature of 5.5$\pm$0.5\,K (see Fig. \ref{fig:CH3CH2CN_Trot}) and a column density of (1.3$\pm$0.1)$\times$10$^{11}$ cm$^{-2}$. 
The abundance ratio of CH$_3$CH$_2$CCH and ethyl cyanide in TMC-1 is 4.8$\pm$0.5.
Although the frequencies of the $^{13}$C, $^{15}$N, and D isotopologs
of ethyl cyanide are well known \citep{Demyk2007,Richard2012,Margules2009}, the intensity of the lines of the main 
isotopolog indicates that the
lines from these isotopologs will be well below the current sensitivity limit of the QUIJOTE line survey.

\begin{table}
\centering
\small
\caption{Derived column densities and abundances}
\label{columndensities}
\centering
\begin{tabular}{{lcccc}}
\hline
Species             & T$_{rot}$  & $N$                        &$X^a$  & \\
                    &  (K)       & (cm$^{-2}$)                &       & \\
\hline                            
CH$_3$CH$_2$CCH     & 9.0        & (6.2$\pm$0.2)$\times$10$^{11}$& 6.2$\times$10$^{-11}$ &A\\
CH$_3$CH$_2$CN      & 5.5$\pm$0.5& (1.3$\pm$0.1)$\times$10$^{11}$& 1.3$\times$10$^{-11}$ & \\
CH$_2$CHCCH         &10.4$\pm$0.9& (9.5$\pm$0.2)$\times$10$^{12}$& 9.5$\times$10$^{-09}$ &B \\
CH$_2$CHCN          & 4.3$\pm$0.2& (6.2$\pm$0.2)$\times$10$^{12}$& 6.2$\times$10$^{-09}$ & \\
$^{13}$CH$_2$CHCN   & 4.3        & (6.2$\pm$0.2)$\times$10$^{10}$& 6.2$\times$10$^{-12}$ &C\\
CH$_2$$^{13}$CHCN   & 4.3        & (5.9$\pm$0.2)$\times$10$^{10}$& 5.9$\times$10$^{-12}$ &C\\
CH$_2$CH$^{13}$CN   & 4.3        & (8.1$\pm$0.3)$\times$10$^{10}$& 8.1$\times$10$^{-12}$ &C\\
CH$_2$CDCN          & 4.3        & (4.7$\pm$0.3)$\times$10$^{10}$& 4.7$\times$10$^{-12}$ &C\\
$trans$-CHDCHCN     & 4.3        & (4.5$\pm$0.3)$\times$10$^{10}$& 4.5$\times$10$^{-12}$ &C\\
$cis$-CHDCHCN       & 4.3        & (4.9$\pm$0.3)$\times$10$^{10}$& 4.9$\times$10$^{-12}$ &C\\
CH$_2$CHC$^{15}$N   & 4.3        & (2.2$\pm$0.2)$\times$10$^{10}$& 2.2$\times$10$^{-12}$ &C\\
\hline
\hline
\end{tabular}
\tablefoot{
\tablefoottext{a}{We assumed $N$(H$_2$)=10$^{22}$ cm$^{-2}$ to derive the molecular abundances
\citep{Cernicharo1987}.}
\tablefoottext{A}{We assumed
that the molecule is thermalized at the kinetic temperature of the cloud (9\,K; see text).}
\tablefoottext{B}{A rotational diagram provides T$_{rot}$=10.5$\pm$1.5. We assumed
that the molecule is thermalized at the kinetic temperature of the cloud (9\,K; see text).}
\tablefoottext{C}{Rotational temperature assumed to be identical to that of the
main isotopolog.}
}
\normalsize
\end{table}

\subsection{CH$_2$CHCCH (vinyl acetylene)}\label{sec:CH2CHCCH}

\begin{figure}
\centering
\includegraphics[scale=0.42]{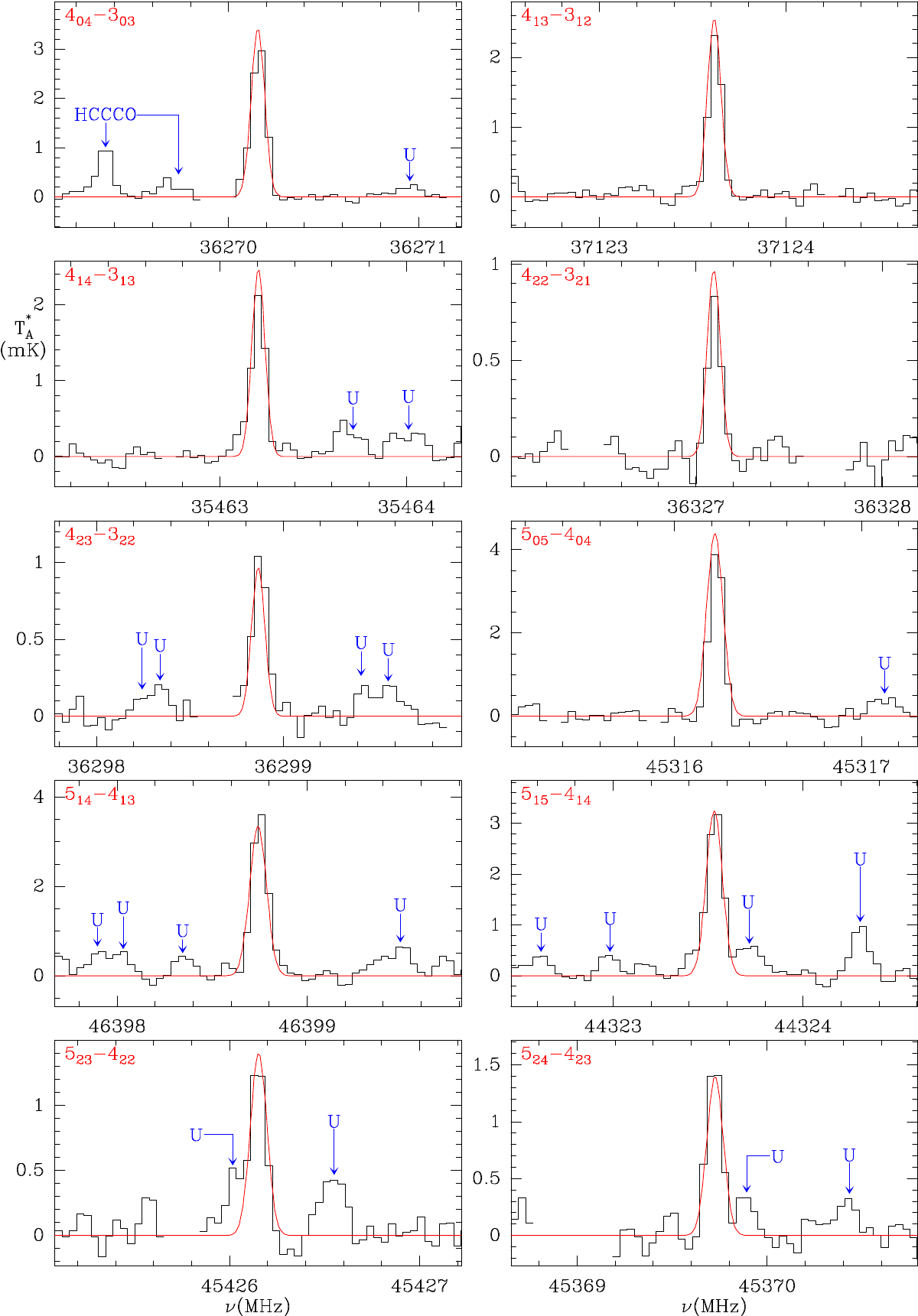}
\caption{Observed lines of CH$_2$CHCCH in TMC-1. 
The line parameters are given in Table \ref{line_parameters}.
The abscissa corresponds to the rest frequency assuming a velocity for the source of 5.83 km\,s$^{-1}$. 
The ordinate is the antenna temperature corrected for atmospheric and telescope losses in mK. 
Quantum numbers are indicated in the top left corner of each panel.
The red line corresponds to the synthetic spectrum derived from the model described in Sect. \ref{sec:CH2CHCCH}. 
Blanked channels correspond to negative features produced in the folding of the frequency-switching data.
}
\label{fig:CH2CHCCH}
\end{figure}

\begin{figure}
\centering
\includegraphics[scale=0.6]{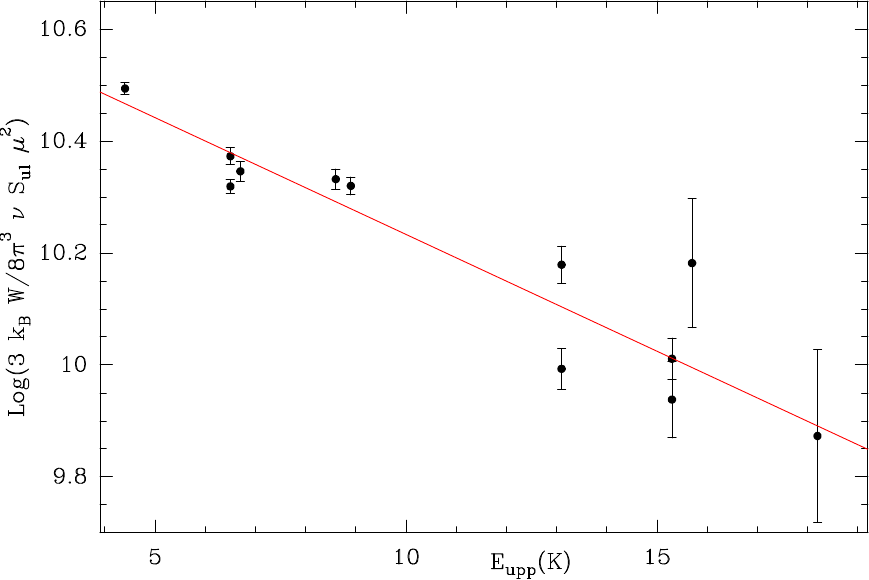}
\caption{Rotational diagram of the observed lines of CH$_2$CHCCH from the data of Table \ref{line_parameters}, including the
lines observed at 3mm.
A rotational temperature of 10.4$\pm$0.9\,K is derived.}
\label{fig:Trot_CH2CHCCH}
\end{figure}

In spite of the similarity in structure with CH$_2$CHCN, which has been detected toward
different astrophysical environments (see Sect. \ref{sec:CH2CHCN}), vinyl acetelyne was
only recently detected 
through the sensitivity of the QUIJOTE line survey \citep{Cernicharo2021b}. Due to the
small dipole moment of the molecule \citep[$\mu_a$=0.43\,D, $\mu_b\sim$0,][]{Sobolev1962,Thorwirth2003}, the derived abundance was rather large ($\sim$10$^{13}$ cm$^{-2}$).
The rotational spectroscopy in the laboratory for this species was summarized by \citet{Thorwirth2003} and \citet{Thorwirth2004}.
The observed lines in our study are shown in Fig. \ref{fig:CH2CHCCH}, and their 
parameters are given in Table \ref{line_parameters}.
A rotational diagram analysis of the data provides T$_{rot}$=10.4$\pm$0.9\,K (see Fig. \ref{fig:Trot_CH2CHCCH}), which is compatible with
a thermalized molecule at the kinetic temperature of the cloud, as expected for a
species with a low dipole moment \citep[$T_K$=9\,K;][]{Agundez2023b}. The derived column density is (9.5$\pm$0.2)$\times$10$^{12}$ cm$^{-2}$, 
which agrees well with the value we reported previously. 
The derived CH$_2$CHCCH/CH$_2$CHCN abundance ratio is 1.5$\pm$0.1. 

\subsection{CH$_2$CHCN (vinyl cyanide; acrylonitrile)}\label{sec:CH2CHCN}
CH$_2$CHCN was detected  in the early years of astrochemistry toward SgrB2 \citep{Gardner1975}. The molecule
was detected toward TMC-1 by \citet{Matthews1981} and more recently by \citet{Cernicharo2021b}.
Vinyl cyanide was also detected toward the carbon-rich star IRC+10216 \citep{Agundez2008}
and in high-mass star-forming regions \citep{Belloche2013,Lopez2014}.
There is a large set of laboratory data on the rotational spectroscopy of this species that
covers frequencies up to 1.67 THz, $J$=129, and $K_a$=28 \citep[][and references therein]{Muller2008,Kisiel2009}.
The frequency predictions have uncertainties lower than 1 kHz in the frequency domains covered in this study.

\begin{figure}
\centering
\includegraphics[scale=0.442]{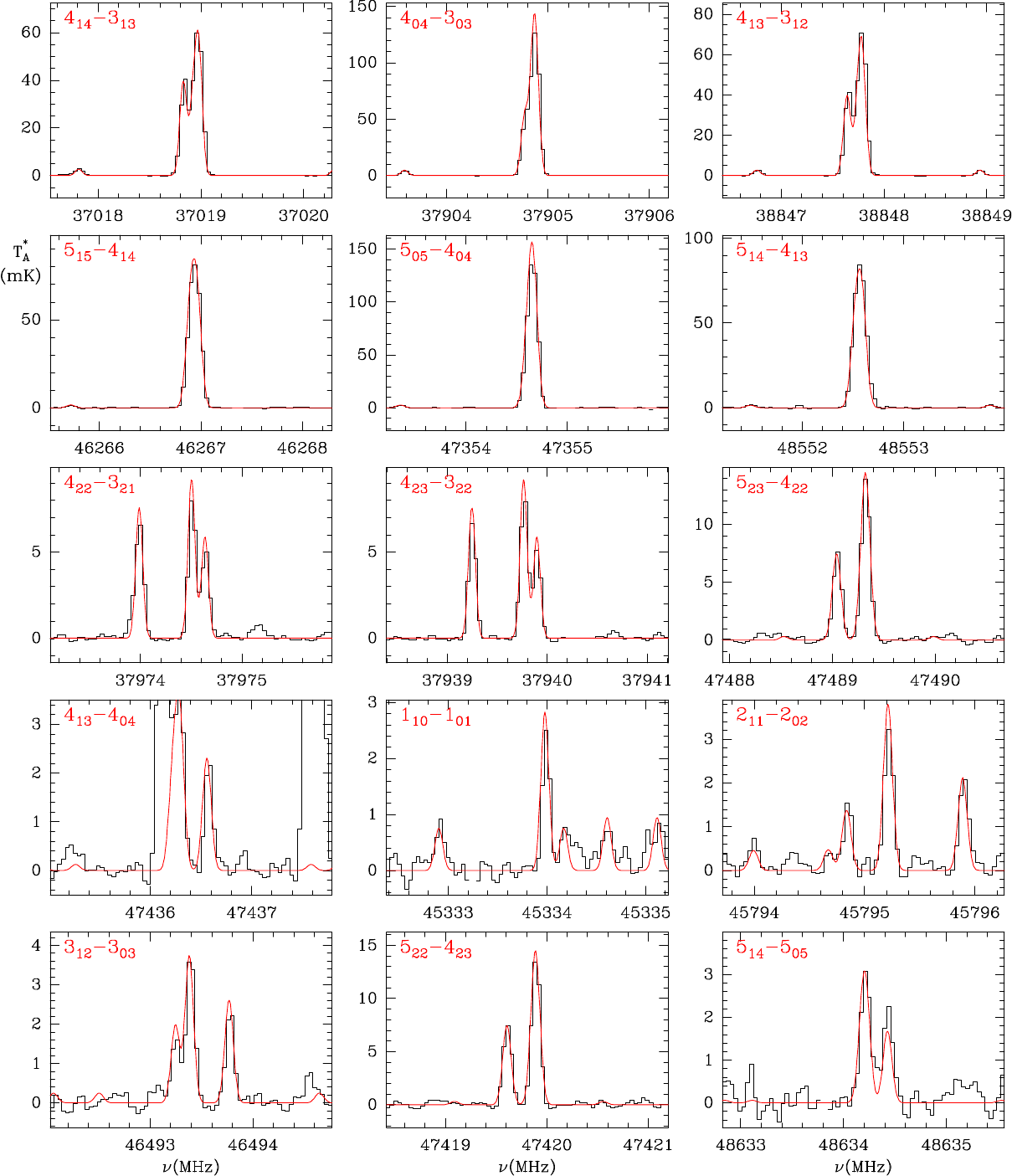}
\caption{Observed lines of CH$_2$CHCN in TMC-1. 
The line parameters are given in Table \ref{line_parameters}.
The abscissa corresponds to the rest frequency assuming a velocity for the source of 5.83 km\,s$^{-1}$. 
The ordinate is the antenna temperature corrected for atmospheric and telescope losses in mK. 
Quantum numbers are indicated in the top left corner of each panel.
The red line corresponds to the synthetic spectrum derived from the model described in Sect. \ref{sec:CH2CHCN}. 
Blanked channels correspond to negative features produced in the folding of the frequency 
switching data. The $J_u$=4,5 ($K_a$=0,1) transitions of the isotopologs of CH$_2$CHCN are detected
and are shown in Fig. \ref{fig:CH2CHCN_iso}.
}
\label{fig:CH2CHCN}
\end{figure}

\begin{figure}
\centering
\includegraphics[scale=0.58]{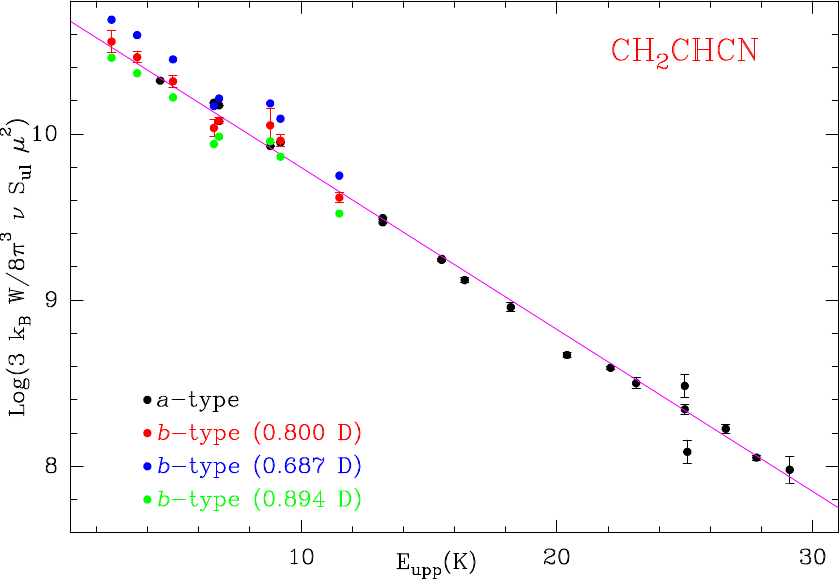}
\caption{Rotational diagram of the observed lines of CH$_2$CHCN. The black and red points correspond
to $a$-type and $b$-type transitions, respectively. A rotational temperature of 4.3$\pm$0.2\,K is
derived. The estimated $b$-component of the dipolar moment is 0.80$\pm$0.03\,D. The blue and green points
correspond to the $b$-type transitions when using $\mu_b$=0.687\,D \citep{Krasnicki2011a} and 0.894\,D \citep{Stolze1985}, respectively.
Their error bars are identical to those of the red points.}
\label{fig:CH2CHCN_Trot}
\end{figure}

We have observed 23 $a$-type and 8 $b$-type transitions of CH$_2$CHCN in the 7\,mm and the 3\,mm domains. Several
of these lines exhibit well-resolved hyperfine components.
The lines observed in the Q band are shown in Fig. \ref{fig:CH2CHCN}. The line parameters for all observed
lines are given in Table \ref{line_parameters}. A rotational diagram is shown in Fig. \ref{fig:CH2CHCN_Trot}.
The initial dipole moments used in the analysis of the data were those measured by \citet{Stolze1985}
($\mu_a$\,=\,3.815\,D and $\mu_b$\,=\,0.894\,D). The $a$- and $b$-type transitions are
well fit with a common rotational temperature of 4.3$\pm$0.2\,K. However, the column density derived
from the $b$-type transitions is a factor 1.3 times lower than the corresponding density from the $a$-type transitions. 
The effect is systematic as all $b$-type transitions are well reproduced with a common column
density. This discrepancy cannot be attributed to a 
calibration problem of the data. All $a$- and $b$-type lines were observed with the
same set of observational data, gathered at the same time, and with the same systematics. 
Moreover, the same upper levels are involved in both transition types. It cannot be
due to an opacity problem because for the strongest $a$-type transitions, we estimate opacities $\sim$0.15. Hence,
it seems that the relative value of the dipole moments is not well determined.

The dipole moment of acrylonitrile has been the subject of several laboratory studies. The first determination
was made by \citet{Wilcox1954}, who derived $\mu_a$\,=\,3.68\,D and $\mu_b$\,=\,1.25\,D. Subsequent 
measurements by \citet{Stolze1985} provided different values, $\mu_a$\,=\,3.815\,$\pm$\,0.012\,D and
$\mu_b$\,=\,0.894\,$\pm$\,0.068\,D. The more recent determination of $\mu_b$\,=\,0.687\,$\pm$\,0.008\,D \citep{Krasnicki2011a} produced the 
reverse, that is, the column density from $b$-type transitions is
1.3 times higher than the value derived from the $a$-type transitions. To coherently
interpret our data, we fit the value of $\mu_b$ necessary to
produce similar column densities for both types of transitions keeping $\mu_a$ to the value
derived by \citet{Stolze1985}. We obtain a value of 0.80$\pm$0.03\,D, which is between the two experimental
determinations. The
rotational diagram shown in Fig. \ref{fig:CH2CHCN_Trot} permits us to simultaneously fit all transitions 
with T$_{rot}$=4.3$\pm$0.2\,K and a column density for acrylonitrile of (6.2$\pm$0.2)$\times$10$^{12}$ cm$^{-2}$
\citep[see also][]{Cernicharo2021b}. The computed synthetic spectra are presented in Fig. \ref{fig:CH2CHCN}
and agree excellently with all the observed transitions, $b$ and $a$ type.

\begin{figure*}
\centering
\includegraphics[scale=0.68,angle=0]{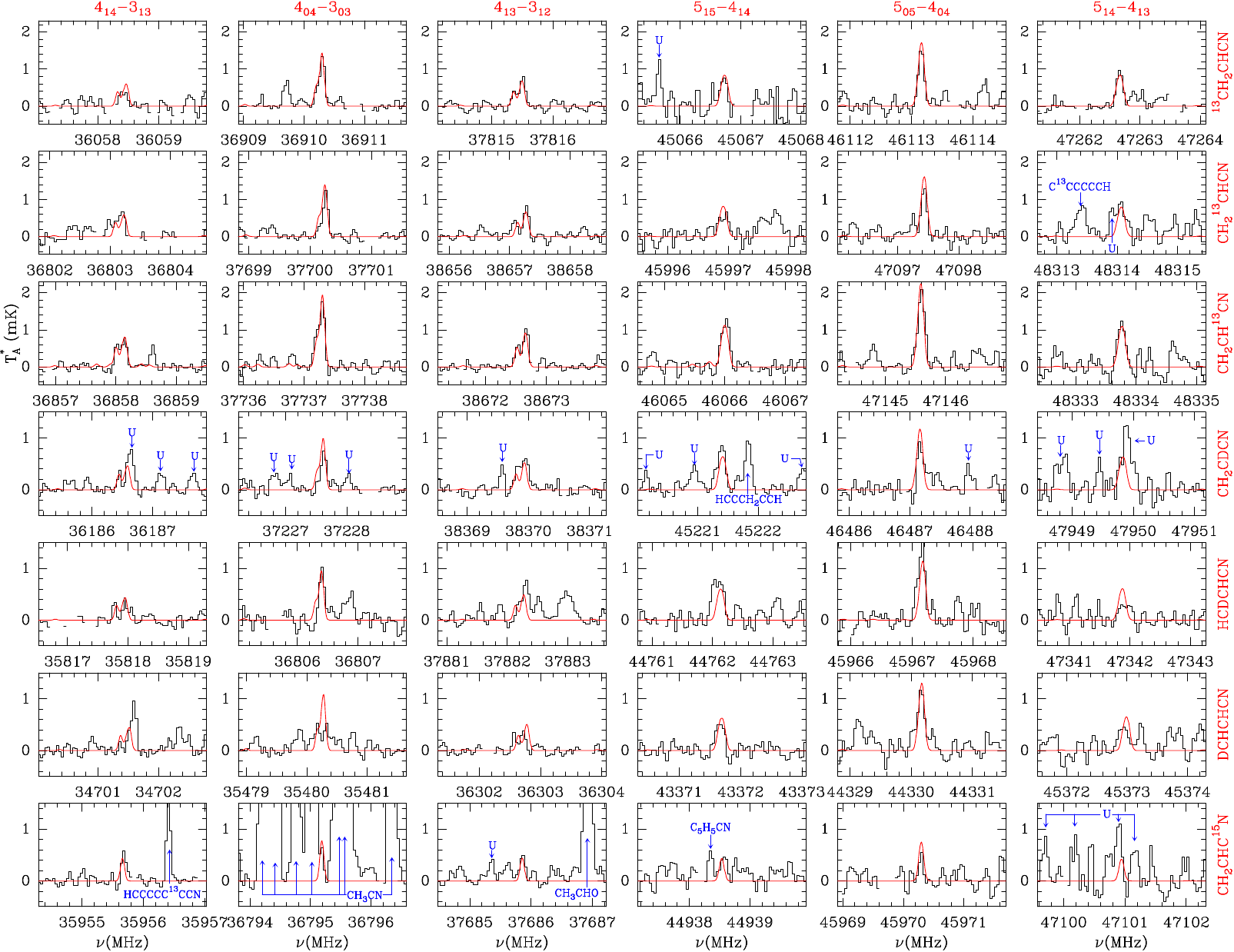}
\caption{Observed lines of the isotopologs of CH$_2$CHCN. 
The line parameters are given in Table \ref{line_parameters}.
The abscissa corresponds to the rest frequency assuming a velocity for the source of 5.83 km\,s$^{-1}$. 
The ordinate is the antenna temperature corrected for atmospheric and telescope losses in mK. 
The species are indicated at the end of each row, and the quantum numbers are indicated at the top of each column.
The red line corresponds to the synthetic spectrum derived from the model described in Sect. \ref{sec:CH2CHCN}. 
Blanked channels correspond to negative features produced in the folding of the frequency-switching data.
}
\label{fig:CH2CHCN_iso}
\end{figure*}

Rotational spectroscopy for the $^{13}$C, $^{15}$N, and deuterated isotopologs of acrylonitrile was
performed by \citet{Colmont1997,Muller2008,Kisiel2009}, and \citet{Krasnicki2011b}. The $^{13}$C isotopologs were detected toward Sgr\,B2 \citep{Muller2008} and Orion-KL \citep{Lopez2014}. 
In our study of TMC-1, we detected six $a$-type transitions for the three $^{13}$C substituted isotopologs, some of which exhibit the corresponding hyperfine splitting. 
We also solidly detected the isotopologs CH$_2$CHC$^{15}$N, 
CH$_2$CDCN, $trans$-CHDCHCN, and $cis$-CHDCHCN (all these lines are $a$ type) for the first time in the ISM. A tentative
detection in Orion-KL of the deuterated species was previously reported \citep{Lopez2014},
but the detection was based on a small number of weak lines with severe
overlap with lines from other molecular species.
The lines observed toward TMC-1 are shown in Fig. \ref{fig:CH2CHCN_iso}. In their
analysis, we adopted the same rotational temperature 
and dipole moments as for the
main isotopolog. The derived column densities for all the isotopologs
are given in Table \ref{columndensities}. The three singly deuterated species show
similar column densities and a deuterium enhancement $\sim$130, which is significantly lower
than was found in other species \citep{Cernicharo2024,Tercero2024}. The 
first and second $^{13}$C isotopologs show a $^{12}$C/$^{13}$C abundance ratio of 100$\pm$7, 
while for the substitution in the third position, the ratio is 76$\pm$5. The
CH$_2$CHCN/CH$_2$CHC$^{15}$N abundance ratio is 282$\pm$35. These values are in line
with those derived for HCCCN by \citet{Tercero2024} and HNCCC and HCCNC by \citet{Cernicharo2024}.
These two papers provide a detailed discussion of the $^{13}$C enhancement in these
species. 

\section{Discussion}\label{sec:discussion}

\begin{figure}
\centering
\includegraphics[width=\columnwidth,angle=0]{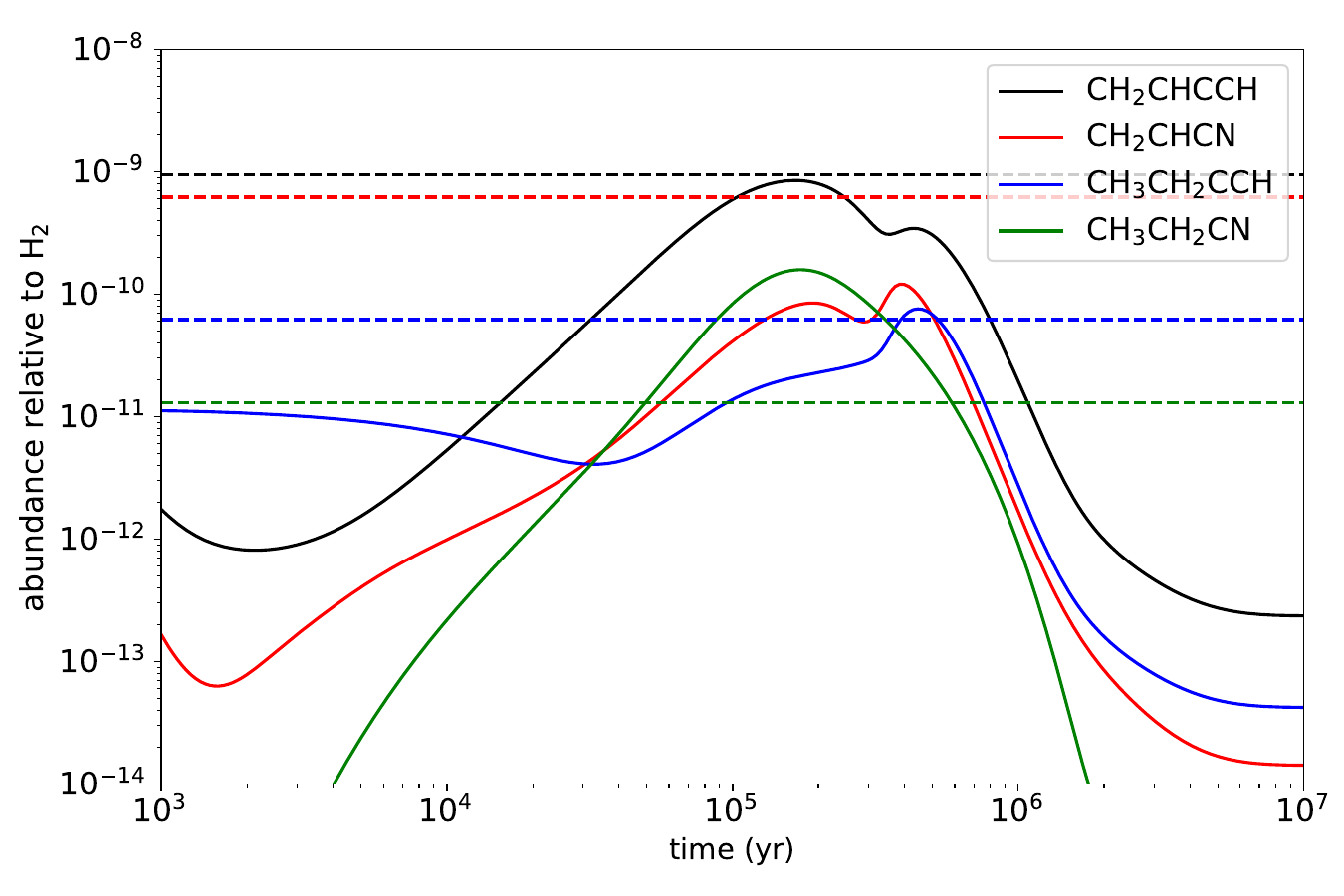}
\caption{Calculated fractional abundances of the CCH and CN derivatives of C$_2$H$_4$ and C$_2$H$_6$ as a function of 
time. The horizontal dashed lines correspond to the abundances observed in TMC-1.}
\label{fig:abun}
\end{figure}

The observational data presented here make it worthwhile to revisit the chemistry of the CCH and CN derivatives of 
CH$_2$CH$_2$ and CH$_3$CH$_3$. We carried out chemical modeling calculations using a gas-phase chemical model. 
The model was essentially the same as presented previously in \cite{Cernicharo2021b}. Briefly, we adopted typical parameters of 
cold dark clouds, that is, a gas kinetic temperature of 10 K, a volume density of H$_2$ of 2\,$\times$\,10$^4$ cm$^{-3}$, 
a cosmic-ray ionization rate of H$_2$ of 1.3\,$\times$\,10$^{-17}$ s$^{-1}$, a visual extinction of 30 mag, and the so-called 
low metal elemental abundances \citep{Agundez2013}. The core of the chemical network is based on the RATE12 network from the 
UMIST database \citep{McElroy2013}, with some updates from the more recent literature (e.g., \citealt{Loison2015}). The relevant 
reactions that describe the formation of the CCH and CN derivatives of CH$_2$CH$_2$ and CH$_3$CH$_3$ are discussed below. 
The calculated fractional abundances of CH$_2$CHCCH, CH$_2$CHCN, 
CH$_3$CH$_2$CCH, and CH$_3$CH$_2$CN are shown as a function of time in Fig.\,\ref{fig:abun}. The chemical model produces 
abundances at times of a few 10$^5$ yr 
that are in reasonable good agreement with the observed values, with discrepancies one order 
of magnitude or smaller. 
Taking into account that the chemistry of these four molecules is not particularly well known, we 
can consider that the agreement between observations and model is satisfactory, although a better knowledge of the 
reactions of formation and destruction of these molecules is highly desirable.

The CCH and CN derivatives of C$_2$H$_4$ are formed in the chemical model through the reactions of C$_2$H and CN with 
C$_2$H$_4$, which are rapid and proceed through H atom elimination, according to an extensive number of studies
\citep{Opansky1996,Vakhtin2001,Bouwman2012,Krishtal2009,Dash2015,Sims1993,Choi2004,Gannon2007,Balucani2015}. 
In addition, CH$_2$CHCCH is also formed by the reaction of CH with CH$_3$CCH and CH$_2$CCH$_2$ \citep{Daugey2005,Goulay2009,Ribeiro2017} 
and by the reaction C$_2$ + C$_2$H$_6$ \citep{Paramo2008}, although this latter reaction probably occurs through H abstraction, in which case, it would not produce CH$_2$CHCCH. On 
the other hand, CH$_2$CHCN is also produced in the reaction CN + CH$_2$CHCH$_3$ \citep{Sims1993,Morales2010,Gannon2007,Huang2009} 
and in the dissociative recombination of the cation C$_3$H$_4$N$^+$, although for this latter process, the branching ratios for 
the different fragmentation products are unknown.

In the case of the CCH and CN derivatives of C$_2$H$_6$, the formation pathways are more uncertain. The most obvious chemical 
route would be CCH + C$_2$H$_6$ and CN + C$_2$H$_6$. These reactions are fast \citep{Opansky1996,Sims1993}, but occur through 
H abstraction rather than H elimination \citep{Dash2015,Georgievskii2007,Espinosa-Garcia2023}. The main routes to CH$_3$CH$_2$CCH 
in the model are the dissociative recombination of C$_4$H$_7^+$ and CH + CH$_2$CHCH$_3$. This latter reaction was measured 
to be rapid at low temperatures \citep{Daugey2005}, and the reaction was found to occur mostly through H elimination 
\citep{Loison2009}. \cite{Trevitt2013} reported that CH$_3$CH$_2$CCH is produced with a branching ratio of 0.12, although theoretical 
calculations \citep{Ribeiro2016,He2019} pointed to 1,2-butadiene as the only C$_4$H$_6$ isomer formed. For CH$_3$CH$_2$CN, the only 
formation channel in the model is the dissociative recombination of C$_3$H$_6$N$^+$, although it is unknown how large cations, 
such as C$_4$H$_7^+$ and C$_3$H$_6$N$^+$, fragment upon reaction with electrons.

In summary, the formation of CH$_2$CHCCH, CH$_2$CHCN, CH$_3$CH$_2$CCH, and CH$_3$CH$_2$CN in TMC-1 can be explained in terms of 
standard gas-phase routes involving neutral-neutral and ion-neutral reactions. However, there are still important uncertainties 
regarding the different products resulting from the various possible reactions of formation of these molecules.

\begin{acknowledgements}
We thank Ministerio de Ciencia e Innovaci\'on of Spain (MICIU) for funding support through projects
PID2019-106110GB-I00, 
and PID2019-106235GB-I00, PID2022-137980NB-I00, MCIN/AEI/ 10.13039/501100011033.
We thank the Consejo Superior de Investigaciones Científicas (CSIC) for funding through project PIE 202250I097.
We also thank ERC for funding through grant ERC-2013-Syg-610256-NANOCOSMOS.

\end{acknowledgements}

\normalsize
\onecolumn
\begin{appendix}

\section{Line parameters}\label{app:lineparameters}
The line parameters for all observed transitions with the Yebes 40m and IRAM 30m radio telescopes
were derived by fitting a Gaussian line profile to them
using the GILDAS package. A
velocity range of $\pm$20\,\kms\, around each feature was considered for the fit after a polynomial 
baseline was removed. Negative features produced in the folding of the frequency switching data were blanked
before baseline removal.

\begin{tiny}
\begin{longtable}{lcccrccrl}
\caption[]{Observed line parameters for the species.
\label{line_parameters}}\\
\hline
\hline
Molecule& $J$$^a$    &  $F$$^b$& $\nu_{rest}$~$^c$ & $\int T_A^* dv$~$^d$& v$_{LSR}$~$^e$ & $\Delta$v~$^f$ & $T_A^*$~$^g$ & Notes \\
        &            &         &  (MHz)            &(mK\,km\,s$^{-1}$)   &(km\,s$^{-1}$)  &(km\,s$^{-1}$)  & (mK)         &       \\
\hline
\endfirsthead
\caption{continued.}\\
\hline
\hline
Molecule& $J$$^a$    &  $F$$^b$& $\nu_{rest}$~$^c$ & $\int T_A^* dv$~$^d$& v$_{LSR}$~$^e$ & $\Delta$v~$^f$ & $T_A^*$~$^g$ & Notes \\
        &            &         &  (MHz)            &(mK\,km\,s$^{-1}$)   &(km\,s$^{-1}$)  &(km\,s$^{-1}$)  & (mK)         &       \\
\hline
\endhead
\hline
\endfoot
\hline
\endlastfoot
\hline
CH$_3$CH$_2$CCH&$4_{1,4}-3_{1,3}$ &           & 33600.696$\pm$0.020&  0.27$\pm$0.05& 5.83$\pm$0.00 & 0.83$\pm$0.18 &  0.31$\pm$0.06 & \\        
               &$4_{0,4}-3_{0,3}$ &           & 34463.749$\pm$0.020&  0.61$\pm$0.07& 5.83$\pm$0.00 & 1.15$\pm$0.15 &  0.50$\pm$0.06 & \\        
               &$4_{2,3}-3_{2,2}$ &           & 34529.087$\pm$0.030&  0.27$\pm$0.09& 5.83$\pm$0.00 & 0.92$\pm$0.25 &  0.28$\pm$0.09 & \\        
               &$4_{1,3}-3_{1,2}$ &           & 35438.395$\pm$0.020&  0.31$\pm$0.07& 5.83$\pm$0.00 & 0.80$\pm$0.00 &  0.36$\pm$0.08 & A\\        
               &$5_{1,5}-4_{1,4}$ &           & 41987.921$\pm$0.030&  0.50$\pm$0.14& 5.83$\pm$0.00 & 1.03$\pm$0.22 &  0.46$\pm$0.19 & B\\        
               &$5_{0,5}-4_{0,4}$ &           & 43027.572$\pm$0.030&  0.62$\pm$0.06& 5.83$\pm$0.00 & 1.00$\pm$0.00 &  0.54$\pm$0.10 & A\\        
               &$5_{2,4}-4_{2,3}$ &           & 43152.200$\pm$0.030&  0.35$\pm$0.09& 5.83$\pm$0.00 & 0.80$\pm$0.00 &  0.43$\pm$0.11 & A\\        
               &$5_{2,3}-4_{2,2}$ &           & 43290.425$\pm$0.030&  0.39$\pm$0.08& 5.83$\pm$0.00 & 0.90$\pm$0.18 &  0.40$\pm$0.08 & \\        
               &$5_{1,4}-4_{1,3}$ &           & 44283.987$\pm$0.030&  0.38$\pm$0.09& 5.83$\pm$0.00 & 0.71$\pm$0.15 &  0.50$\pm$0.09 & \\   
\\
CH$_3$CH$_2$CN &$4_{1,4}-3_{1,3}$ &  4-3      & 34823.952$\pm$0.001&  0.90$\pm$0.09& 5.57$\pm$0.11 & 0.80$\pm$0.00 &  1.06$\pm$0.06 & C\\
               &                &5-4 \& 3-2   & 34824.090$\pm$0.001&  1.02$\pm$0.09& 5.84$\pm$0.10 & 0.80$\pm$0.00 &  1.19$\pm$0.06 & C,D\\
               &$4_{0,4}-3_{0,3}$ & 3-2       & 35722.142$\pm$0.001&  0.95$\pm$0.06& 5.65$\pm$0.25 & 0.80$\pm$0.00 &  1.11$\pm$0.08 & C\\
               &                &5-4 \& 4-3   & 35722.239$\pm$0.001&  1.72$\pm$0.06& 5.81$\pm$0.09 & 0.70$\pm$0.10 &  2.38$\pm$0.08 & C,D\\
               &$4_{2,3}-3_{2,2}$ & 4-3       & 35791.730$\pm$0.002&  0.20$\pm$0.06& 5.83$\pm$0.11 & 0.69$\pm$0.19 &  0.27$\pm$0.08 & \\
               &                & 5-4         & 35792.188$\pm$0.001&  0.33$\pm$0.07& 5.81$\pm$0.08 & 0.70$\pm$0.16 &  0.44$\pm$0.08 & \\
               &                & 3-2         & 35792.305$\pm$0.001&  0.22$\pm$0.06& 5.85$\pm$0.08 & 0.52$\pm$0.22 &  0.39$\pm$0.08 & \\
               &$4_{2,2}-3_{2,1}$ & 4-3       & 35865.827$\pm$0.002&  0.32$\pm$0.06& 6.06$\pm$0.11 & 0.57$\pm$0.14 &  0.52$\pm$0.10 & E\\
               &                & 5-4         & 35866.279$\pm$0.001&  0.25$\pm$0.08& 5.72$\pm$0.08 & 0.56$\pm$0.17 &  0.42$\pm$0.10 & E\\
               &                & 3-2         & 35866.396$\pm$0.001&  0.33$\pm$0.09& 5.82$\pm$0.08 & 0.73$\pm$0.25 &  0.42$\pm$0.10 & E\\
               &$4_{1,3}-3_{1,2}$ & 3-2       & 36379.581$\pm$0.001&  0.69$\pm$0.09& 5.71$\pm$0.08 & 0.78$\pm$0.12 &  0.84$\pm$0.08 & C\\
               &                &5-4 \& 4-3   & 36379.712$\pm$0.001&  1.44$\pm$0.09& 5.78$\pm$0.00 & 0.74$\pm$0.05 &  1.82$\pm$0.08 & C,D\\
               &$5_{1,5}-4_{1,3}$ & Main      & 43516.229$\pm$0.001&  1.76$\pm$0.12& 5.77$\pm$0.03 & 0.89$\pm$0.08 &  1.86$\pm$0.12 & D\\
               &$5_{0,5}-4_{0,4}$ & Main      & 44597.010$\pm$0.001&  2.86$\pm$0.14& 5.93$\pm$0.02 & 0.88$\pm$0.05 &  3.05$\pm$0.11 & D\\
               &$5_{2,3}-4_{2,2}$ & 5-4       & 44877.939$\pm$0.001&  0.71$\pm$0.06& 6.13$\pm$0.08 & 1.00$\pm$0.22 &  0.67$\pm$0.13 & \\
               &                &6-5 \& 4-3   & 44878.172$\pm$0.001&  0.73$\pm$0.08& 5.97$\pm$0.08 & 1.11$\pm$0.21 &  0.62$\pm$0.13 & D\\
               &$5_{1,4}-4_{1,3}$ & Main      & 45909.544$\pm$0.001&  1.89$\pm$0.12& 5.93$\pm$0.02 & 0.78$\pm$0.06 &  2.28$\pm$0.10 & \\
\\
CH$_2$CHCCH    &$4_{1,4}-3_{1,3}$ &           & 35463.207$\pm$0.001&  1.76$\pm$0.06& 5.85$\pm$0.01 & 0.79$\pm$0.03 &  2.10$\pm$0.07 & \\
               &$4_{0,4}-3_{1,0}$ &           & 36270.155$\pm$0.001&  2.53$\pm$0.06& 5.80$\pm$0.01 & 0.77$\pm$0.02 &  3.10$\pm$0.06 & \\
               &$4_{2,3}-3_{2,2}$ &           & 36298.867$\pm$0.001&  0.92$\pm$0.07& 5.77$\pm$0.03 & 0.83$\pm$0.07 &  1.04$\pm$0.08 & \\
               &$4_{2,2}-3_{2,1}$ &           & 36327.096$\pm$0.001&  0.60$\pm$0.05& 5.80$\pm$0.03 & 0.69$\pm$0.07 &  0.83$\pm$0.08 & \\
               &$4_{1,3}-3_{1,2}$ &           & 37123.615$\pm$0.001&  1.72$\pm$0.07& 5.79$\pm$0.03 & 0.69$\pm$0.03 &  2.36$\pm$0.06 & \\
               &$5_{1,5}-4_{1,4}$ &           & 44323.533$\pm$0.001&  2.45$\pm$0.10& 5.82$\pm$0.01 & 0.71$\pm$0.04 &  3.25$\pm$0.11 & \\
               &$5_{0,5}-4_{0,4}$ &           & 45316.216$\pm$0.001&  2.54$\pm$0.07& 5.78$\pm$0.01 & 0.57$\pm$0.03 &  4.19$\pm$0.12 & \\
               &$5_{2,4}-4_{2,3}$ &           & 45369.726$\pm$0.001&  1.05$\pm$0.09& 5.86$\pm$0.03 & 0.63$\pm$0.06 &  1.58$\pm$0.12 & \\
               &$5_{2,3}-4_{2,2}$ &           & 45426.153$\pm$0.001&  0.89$\pm$0.14& 5.85$\pm$0.05 & 0.61$\pm$0.10 &  1.37$\pm$0.10 & \\
               &$5_{1,4}-4_{1,3}$ &           & 46398.742$\pm$0.001&  2.49$\pm$0.09& 5.77$\pm$0.01 & 0.63$\pm$0.03 &  3.70$\pm$0.13 & \\
               &$8_{0,8}-7_{0,7}$ &           & 72357.946$\pm$0.001&  8.39$\pm$2.23& 5.90$\pm$0.08 & 0.58$\pm$0.16 & 13.69$\pm$4.21 & \\
               &$8_{1,7}-7_{1,6}$ &           & 74197.205$\pm$0.001&  4.17$\pm$1.48& 5.54$\pm$0.07 & 0.34$\pm$0.12 & 11.41$\pm$3.69 & \\%919-818 fully blended ith o-CH2CN
               &$9_{0,9}-8_{0,8}$ &           & 81330.920$\pm$0.001&  4.56$\pm$0.44& 5.78$\pm$0.03 & 0.66$\pm$0.08 &  6.52$\pm$0.81 & \\
               &$9_{1,8}-8_{1,7}$ &           & 83451.437$\pm$0.001&  3.63$\pm$0.33& 5.89$\pm$0.03 & 0.61$\pm$0.07 &  5.56$\pm$0.67 & \\
               &$10_{1,10}-9_{1,9}$&          & 88558.997$\pm$0.001&               &               &               &  $\le$7.53     & \\
               &$10_{0,10}-9_{0,9}$&          & 90279.458$\pm$0.001&               &               &               &  $\le$7.59     & \\
               &$10_{1,9}-9_{1,8}$&           & 92698.125$\pm$0.001&  1.46$\pm$0.33& 6.19$\pm$0.10 & 0.61$\pm$0.13 &  2.23$\pm$0.84 &F\\

\\
CH$_2$CHCN     &$4_{1,4}-3_{1,3}$ & 4-4       & 37017.811$\pm$0.001&  2.42$\pm$0.11& 5.80$\pm$0.02 & 0.78$\pm$0.05 &  2.90$\pm$0.10 & \\
               &                & 4-3         & 37018.828$\pm$0.001& 30.87$\pm$0.15& 5.74$\pm$0.01 & 0.73$\pm$0.01 & 39.86$\pm$0.10 & \\          
               &                &5-4 \&3-2    & 37018.981$\pm$0.001& 55.77$\pm$0.15& 5.91$\pm$0.01 & 0.84$\pm$0.01 & 62.76$\pm$0.10 &D\\          
               &                & 3-3         & 37020.299$\pm$0.001&  1.73$\pm$0.09& 5.80$\pm$0.02 & 0.64$\pm$0.04 &  2.56$\pm$0.10 & \\
               &$4_{0,4}-3_{0,3}$ &   4-4     & 37903.585$\pm$0.001&  3.21$\pm$0.30& 5.80$\pm$0.02 & 0.68$\pm$0.07 &  4.42$\pm$0.11& \\
               &                &   3-2       & 37904.770$\pm$0.001& 39.15$\pm$0.37& 5.73$\pm$0.03 & 0.75$\pm$0.03 & 48.98$\pm$0.11& \\
               &                &5-4 \& 4-3   & 37904.880$\pm$0.001& 95.35$\pm$0.38& 5.85$\pm$0.03 & 0.72$\pm$0.03 &124.52$\pm$0.11&D\\
               &                &   3-3       & 37906.477$\pm$0.001&  1.93$\pm$0.06& 5.78$\pm$0.02 & 0.58$\pm$0.02 &  3.15$\pm$0.11&G\\
               &$4_{2,3}-3_{2,2}$ &4-4 \& 4-3 & 37939.247$\pm$0.001&  4.87$\pm$0.06& 5.78$\pm$0.01 & 0.67$\pm$0.01 &  6.80$\pm$0.08 & \\
               &                &  5-4        & 37939.764$\pm$0.001&  6.25$\pm$0.07& 5.79$\pm$0.01 & 0.71$\pm$0.01 &  8.23$\pm$0.08 &D\\
               &                &3-2 \& 3-3   & 37939.897$\pm$0.001&  3.60$\pm$0.07& 5.79$\pm$0.01 & 0.66$\pm$0.02 &  5.17$\pm$0.08 &D\\
               &$4_{2,2}-3_{2,1}$ &4-3 \& 4-4 & 37973.988$\pm$0.001&  5.73$\pm$0.08& 5.82$\pm$0.01 & 0.81$\pm$0.01 &  6.67$\pm$0.10 &D\\
               &                &   5-4       & 37974.504$\pm$0.001&  5.97$\pm$0.09& 5.79$\pm$0.01 & 0.68$\pm$0.01 &  8.30$\pm$0.10 &\\
               &                &3-2 \& 3-3   & 37974.634$\pm$0.001&  3.92$\pm$0.09& 5.76$\pm$0.01 & 0.71$\pm$0.02 &  5.18$\pm$0.10 &D\\
               &$4_{1,3}-3_{1,2}$ &  4-4      & 38846.763$\pm$0.001&  2.03$\pm$0.08& 5.76$\pm$0.01 & 0.71$\pm$0.03 &  2.70$\pm$0.10& \\
               &                &  4-3        & 38847.641$\pm$0.001& 31.24$\pm$0.12& 5.73$\pm$0.02 & 0.70$\pm$0.01 & 41.80$\pm$0.10& \\
               &                &5-4 \& 3-2   & 38847.790$\pm$0.001& 59.73$\pm$0.12& 5.87$\pm$0.02 & 0.78$\pm$0.01 & 72.44$\pm$0.10&D\\
               &                &  3-3        & 38848.933$\pm$0.001&  2.07$\pm$0.07& 5.78$\pm$0.01 & 0.64$\pm$0.03 &  3.06$\pm$0.10& \\
               &$1_{1,0}-1_{0,1}$ & 1-0       & 45332.905$\pm$0.001&  0.75$\pm$0.12& 5.74$\pm$0.06 & 0.82$\pm$0.18 &  0.81$\pm$0.11& \\ 
               &                & 2-2         & 45333.979$\pm$0.001&  1.53$\pm$0.10& 5.75$\pm$0.02 & 0.56$\pm$0.04 &  2.54$\pm$0.11& \\ 
               &                & 0-1         & 45334.170$\pm$0.001&  0.48$\pm$0.12& 5.73$\pm$0.14 & 0.72$\pm$0.08 &  0.65$\pm$0.11&H\\ 
               &                & 1-2         & 45334.610$\pm$0.001&  0.79$\pm$0.15& 5.81$\pm$0.11 & 1.09$\pm$0.20 &  0.68$\pm$0.11& \\ 
               &                & 2-1         & 45335.116$\pm$0.001&  0.60$\pm$0.13& 5.64$\pm$0.07 & 0.65$\pm$0.16 &  0.87$\pm$0.11& \\ 
               &                & 1-1         & 45335.747$\pm$0.001&  0.41$\pm$0.09& 5.61$\pm$0.08 & 0.58$\pm$0.09 &  0.67$\pm$0.11& \\ 
               &$2_{1,1}-2_{0,2}$ & 2-1       & 45793.994$\pm$0.001&  0.50$\pm$0.06& 5.81$\pm$0.04 & 0.65$\pm$0.09 &  0.72$\pm$0.10& \\
               &                & 2-3         & 45794.671$\pm$0.001&  0.35$\pm$0.07& 5.60$\pm$0.10 & 0.84$\pm$0.20 &  0.39$\pm$0.10& \\
               &                & 1-1         & 45794.837$\pm$0.001&  0.92$\pm$0.07& 5.74$\pm$0.02 & 0.53$\pm$0.04 &  1.63$\pm$0.10& \\
               &                & 3-3         & 45795.213$\pm$0.001&  2.20$\pm$0.06& 5.78$\pm$0.02 & 0.63$\pm$0.02 &  3.27$\pm$0.10& \\
               &                & 2-2         & 45795.889$\pm$0.001&  1.53$\pm$0.06& 5.76$\pm$0.02 & 0.68$\pm$0.03 &  2.12$\pm$0.10& \\
               &                & 3-2         & 45796.431$\pm$0.001&  0.49$\pm$0.08& 5.70$\pm$0.09 & 1.11$\pm$0.21 &  0.42$\pm$0.10& \\
               &                & 1-2         & 45796.732$\pm$0.001&  0.18$\pm$0.05& 5.99$\pm$0.08 & 0.53$\pm$0.14 &  0.32$\pm$0.10& \\               
               &$5_{1,5}-4_{1,4}$ & 5-5       & 46265.718$\pm$0.001&  0.89$\pm$0.09& 5.87$\pm$0.03 & 0.56$\pm$0.05 &  1.50$\pm$0.14& \\
               &                & Main        & 46266.971$\pm$0.001& 77.60$\pm$0.11& 6.05$\pm$0.01 & 0.87$\pm$0.01 & 83.80$\pm$0.14&D\\
               &                & 4-4         & 46268.396$\pm$0.001&  0.87$\pm$0.08& 5.78$\pm$0.03 & 0.55$\pm$0.06 &  1.50$\pm$0.14& \\
               &$3_{1,2}-3_{0,3}$ & 2-2       & 46493.246$\pm$0.001&  0.81$\pm$0.11& 5.77$\pm$0.03 & 0.55$\pm$0.07 &  1.37$\pm$0.15& \\
               &                & 4-4         & 46493.381$\pm$0.001&  2.43$\pm$0.06& 5.73$\pm$0.01 & 0.61$\pm$0.02 &  3.78$\pm$0.15& \\
               &                & 3-3         & 46493.768$\pm$0.001&  1.53$\pm$0.04& 5.75$\pm$0.01 & 0.59$\pm$0.03 &  2.47$\pm$0.15& \\
               &                & 4-3         & 46494.646$\pm$0.001&  0.52$\pm$0.12& 6.10$\pm$0.13 & 1.12$\pm$0.22 &  0.44$\pm$0.15&H\\
               &                & 2-3         & 46494.953$\pm$0.001&  0.87$\pm$0.20& 5.79$\pm$0.20 & 1.00$\pm$0.25 &  0.82$\pm$0.15&H\\
               &$5_{0,5}-4_{0,4}$ & 5-5       & 47353.354$\pm$0.001&  1.67$\pm$0.09& 5.84$\pm$0.02 & 0.60$\pm$0.04 &  2.64$\pm$0.15& \\
               &                & Main        & 47354.670$\pm$0.001&112.87$\pm$0.10& 5.94$\pm$0.01 & 0.75$\pm$0.01 &142.33$\pm$0.15&D\\
               &                & 4-4         & 47356.233$\pm$0.001&  1.52$\pm$0.09& 5.87$\pm$0.02 & 0.61$\pm$0.04 &  2.34$\pm$0.15& \\
               &$5_{2,4}-4_{2,3}$ & 5-4       & 47419.606$\pm$0.001&  5.03$\pm$0.13& 5.79$\pm$0.01 & 0.62$\pm$0.02 &  7.58$\pm$0.18& \\
               &                &6-5 \& 4-3   & 47419.877$\pm$0.001&  9.22$\pm$0.13& 5.74$\pm$0.01 & 0.62$\pm$0.02 & 14.05$\pm$0.18&D\\
               &$4_{1,3}-4_{0,4}$ &5-5 \& 3-3 & 47436.292$\pm$0.001&  3.06$\pm$0.11& 5.89$\pm$0.02 & 0.74$\pm$0.04 &  3.86$\pm$0.10&D\\
               &                & 4-4         & 47436.559$\pm$0.001&  1.43$\pm$0.08& 5.72$\pm$0.02 & 0.58$\pm$0.04 &  2.32$\pm$0.10& \\
               &$5_{2,3}-4_{2,2}$ & 5-4       & 47489.042$\pm$0.001&  5.01$\pm$0.04& 5.80$\pm$0.01 & 0.62$\pm$0.01 &  7.59$\pm$0.18& \\
               &                &6-5 \& 4-3   & 47489.311$\pm$0.001&  9.44$\pm$0.04& 5.73$\pm$0.01 & 0.63$\pm$0.01 & 14.13$\pm$0.18&D\\
               &$5_{1,4}-4_{1,3}$ & 4-5       & 48551.488$\pm$0.001&  1.26$\pm$0.16& 5.77$\pm$0.04 & 0.60$\pm$0.08 &  1.99$\pm$0.22& \\
               &                &Main         & 48552.598$\pm$0.001& 79.78$\pm$0.18& 6.01$\pm$0.01 & 0.87$\pm$0.01 & 86.41$\pm$0.22&D\\
               &                &4-4          & 48553.851$\pm$0.001&  1.24$\pm$0.17& 5.84$\pm$0.04 & 0.58$\pm$0.09 &  2.03$\pm$0.22& \\
               &$5_{1,4}-5_{0,5}$ &4-4 \& 6-6 & 48634.220$\pm$0.001&  2.39$\pm$0.17& 5.84$\pm$0.03 & 0.74$\pm$0.07 &  3.05$\pm$0.19&D\\             
               &                & 5-5         & 48634.427$\pm$0.001&  1.70$\pm$0.17& 5.79$\pm$0.04 & 0.73$\pm$0.09 &  2.19$\pm$0.19& \\ 
               &$8_{1,8}-7_{1,7}$ & Main      & 73981.555$\pm$0.001& 39.79$\pm$2.58& 5.80$\pm$0.02 & 0.53$\pm$0.04 & 70.52$\pm$5.36& \\
               &$8_{0,8}-7_{0,7}$ & Main      & 75585.693$\pm$0.001& 60.11$\pm$1.90& 5.78$\pm$0.01 & 0.55$\pm$0.02 &103.47$\pm$3.93& \\
               &$8_{2,7}-7_{2,6}$ & Main      & 75838.862$\pm$0.001& 13.07$\pm$2.11& 5.79$\pm$0.04 & 0.62$\pm$0.14 & 19.67$\pm$2.89&H\\
               &$8_{2,6}-7_{2,5}$ & Main      & 76128.883$\pm$0.001&  5.25$\pm$0.85& 5.82$\pm$0.03 & 0.38$\pm$0.06 & 13.14$\pm$2.07& \\
               &$4_{1,4}-3_{0,3}$ &   3-2     & 80768.685$\pm$0.001&  2.30$\pm$0.42& 5.83$\pm$0.01 & 0.56$\pm$0.09 &  3.86$\pm$0.51& \\
               &                  &5-4 \& 4-3 & 80768.827$\pm$0.001&  5.04$\pm$0.44& 6.36$\pm$0.05 & 0.65$\pm$0.06 &  7.28$\pm$0.51& \\
               &$9_{1,9}-8_{1,8}$  & Main     & 83207.507$\pm$0.001& 21.83$\pm$0.34& 5.80$\pm$0.01 & 0.57$\pm$0.01 & 35.96$\pm$0.70& \\
               &$9_{0,9}-8_{0,8}$  & Main     & 84946.003$\pm$0.001& 26.90$\pm$0.85& 5.81$\pm$0.01 & 0.50$\pm$0.02 & 60.18$\pm$1.26& \\
               &$9_{2,8}-8_{2,7}$  & Main     & 85302.649$\pm$0.001&  5.23$\pm$0.95& 5.73$\pm$0.03 & 0.35$\pm$0.07 & 13.94$\pm$2.51& \\
               &$9_{2,7}-8_{2,6}$  & Main     & 85715.426$\pm$0.001& 15.79$\pm$3.37& 5.75$\pm$0.12 & 1.22$\pm$0.22 & 12.12$\pm$2.92&H\\
               &$9_{1,8}-8_{1,7}$  & Main     & 83712.818$\pm$0.001& 18.47$\pm$1.40& 5.82$\pm$0.02 & 0.50$\pm$0.04 & 34.86$\pm$3.26& \\
               &$5_{1,4}-4_{0,4}$  & 4-3      & 89130.841$\pm$0.001&  7.12$\pm$1.40& 5.70$\pm$0.12 & 1.42$\pm$0.04 &  4.72$\pm$3.26&H \\
               &$5_{1,4}-4_{0,4}$  &6-5 \& 5-4& 89130.919$\pm$0.001&  2.99$\pm$0.99& 5.76$\pm$0.03 & 0.37$\pm$0.08 &  7.69$\pm$1.37&D,H\\
               &$10_{1,10}-9_{1,9}$& Main     & 92426.251$\pm$0.001& 11.50$\pm$0.73& 5.80$\pm$0.02 & 0.49$\pm$0.03 & 21.94$\pm$1.72& \\
               &$10_{0,10}-9_{0,9}$& Main     & 94276.637$\pm$0.001& 15.45$\pm$1.06& 5.81$\pm$0.02 & 0.61$\pm$0.05 & 23.89$\pm$1.98& \\
               &$10_{2,9}-9_{2,8}$ & Main     & 94760.783$\pm$0.001&  3.59$\pm$0.59& 5.82$\pm$0.04 & 0.49$\pm$0.10 &  6.96$\pm$1.23& \\
               &$10_{2,8}-9_{2,7}$ & Main     & 95325.476$\pm$0.001&  2.72$\pm$0.44& 5.91$\pm$0.06 & 0.74$\pm$0.16 &  3.45$\pm$0.68& \\
               &$10_{1,9}-9_{1,8}$ & Main     & 96982.440$\pm$0.001&  8.04$\pm$0.28& 5.80$\pm$0.01 & 0.54$\pm$0.02 & 13.96$\pm$0.56& \\
               &$6_{1,6}-5_{0,5}$  & Main     & 97286.826$\pm$0.001&  4.73$\pm$0.35& 5.80$\pm$0.02 & 0.60$\pm$0.05 &  7.44$\pm$0.66& \\
%               &$3_{2,1}-4_{1,4}$  & Main     & 99659.307$\pm$0.001&  1.58$\pm$0.47& 5.81$\pm$0.09 & 0.55$\pm$0.20 &  2.72$\pm$0.76&I \\
             &$11_{1,11}-10_{1,10}$& Main     &101637.233$\pm$0.001&  3.66$\pm$0.94& 6.01$\pm$0.06 & 0.39$\pm$0.14 &  8.84$\pm$2.03&F\\
             &$11_{0,11}-10_{0,10}$& Main     &103575.398$\pm$0.001&  4.31$\pm$1.54& 6.03$\pm$0.11 & 0.54$\pm$0.18 &  7.45$\pm$3.97&F\\
             &$11_{1,10}-10_{1,9}$ & Main     &106641.391$\pm$0.001&  8.18$\pm$1.71& 5.67$\pm$0.08 & 0.73$\pm$0.18 & 10.39$\pm$3.53&F\\
\\
$^{13}$CH$_2$CHCN     &$4_{1,4}-3_{1,3}$ & 4-3        & 36058.327$\pm$0.001&  0.20$\pm$0.09& 5.47$\pm$0.11 & 0.47$\pm$0.42 &  0.40$\pm$0.20 & F\\
                      &$4_{1,4}-3_{1,3}$ & 3-2 \& 5-4 & 36058.479$\pm$0.001&  0.23$\pm$0.32& 5.97$\pm$0.43 & 0.58$\pm$0.66 &  0.37$\pm$0.20 & D\\
                      &$4_{0,4}-3_{0,3}$ & 3-2        & 36910.202$\pm$0.001&  0.33$\pm$0.28& 5.72$\pm$0.27 & 0.64$\pm$0.20 &  0.48$\pm$0.12 & \\
                      &$4_{0,4}-3_{0,3}$ & 4-3 \& 5-4  & 36910.311$\pm$0.001&  1.06$\pm$0.30& 5.82$\pm$0.10 & 0.72$\pm$0.20 &  1.39$\pm$0.12 & D\\
                      &$4_{1,3}-3_{1,2}$ & 4-3        & 37815.361$\pm$0.001&  0.30$\pm$0.09& 5.76$\pm$0.10 & 0.70$\pm$0.21 &  0.40$\pm$0.10 & \\ 
                      &$4_{1,3}-3_{1,2}$ & 3-2 \& 5-4 & 37815.510$\pm$0.001&  0.65$\pm$0.10& 5.86$\pm$0.05 & 0.76$\pm$0.08 &  0.81$\pm$0.10 & D\\ 
                      &$5_{1,5}-4_{1,4}$ & Main       & 45066.772$\pm$0.001&  0.70$\pm$0.24& 6.21$\pm$0.15 & 0.82$\pm$0.33 &  0.80$\pm$0.34 & \\    
                      &$5_{0,5}-4_{0,4}$ & Main       & 46113.185$\pm$0.001&  1.21$\pm$0.11& 5.97$\pm$0.03 & 0.73$\pm$0.08 &  1.57$\pm$0.18 & \\   
                      &$5_{1,4}-4_{1,3}$ & Main       & 47262.702$\pm$0.001&  0.81$\pm$0.13& 6.10$\pm$0.06 & 0.76$\pm$0.14 &  1.00$\pm$0.21 & \\    
\\
CH$_2$$^{13}$CHCN     &$4_{1,4}-3_{1,3}$ & 4-3        & 36803.089$\pm$0.001&  0.56$\pm$0.53& 5.58$\pm$0.52 & 1.17$\pm$1.12 &  0.45$\pm$0.15 & F, H\\
                      &$4_{1,4}-3_{1,3}$ & 3-2 \& 5-4 & 36803.242$\pm$0.001&  0.36$\pm$0.28& 6.03$\pm$0.14 & 0.62$\pm$0.31 &  0.55$\pm$0.15 & D, F\\
                      &$4_{0,4}-3_{0,3}$ & 3-2        & 37700.144$\pm$0.001&  0.24$\pm$0.23& 5.71$\pm$0.40 & 0.76$\pm$0.59 &  0.29$\pm$0.12 & \\         
                      &$4_{0,4}-3_{0,3}$ & 4-3 \& 5-4  & 37700.255$\pm$0.001&  0.86$\pm$0.24& 5.80$\pm$0.08 & 0.65$\pm$0.13 &  1.24$\pm$0.12 & D\\ 
                      &$4_{1,3}-3_{1,2}$ & 4-3        & 38657.133$\pm$0.001&  0.41$\pm$0.06& 6.23$\pm$0.07 & 0.88$\pm$0.18 &  0.43$\pm$0.08 & H\\ 
                      &$4_{1,3}-3_{1,2}$ & 3-2 \& 5-4 & 38657.283$\pm$0.001&  0.66$\pm$0.05& 5.88$\pm$0.02 & 0.71$\pm$0.03 &  0.87$\pm$0.08 & D\\ 
                      &$5_{1,5}-4_{1,4}$ & Main       & 45996.957$\pm$0.001&  0.93$\pm$0.18& 6.14$\pm$0.16 & 1.66$\pm$0.40 &  0.53$\pm$0.18 & F\\
                      &$5_{0,5}-4_{0,4}$ & Main       & 47097.459$\pm$0.001&  0.94$\pm$0.10& 5.97$\pm$0.03 & 0.65$\pm$0.07 &  1.37$\pm$0.16 & \\   
                      &$5_{1,4}-4_{1,3}$ & Main       & 48314.102$\pm$0.001&  1.17$\pm$0.28& 6.13$\pm$0.16 & 1.27$\pm$0.35 &  0.87$\pm$0.25 & H\\    
\\
CH$_2$CH$^{13}$CN     &$4_{1,4}-3_{1,3}$ & 4-3        & 36858.005$\pm$0.001&  0.46$\pm$0.09& 5.70$\pm$0.07 & 0.65$\pm$0.13 &  0.67$\pm$0.08 & \\
                      &$4_{1,4}-3_{1,3}$ & 3-2 \& 5-4 & 36858.158$\pm$0.001&  0.70$\pm$0.10& 5.95$\pm$0.06 & 0.82$\pm$0.13 &  0.81$\pm$0.08 & D\\
                      &$4_{0,4}-3_{0,3}$ & 3-2        & 37737.209$\pm$0.001&  1.08$\pm$0.30& 5.68$\pm$0.14 & 0.97$\pm$0.23 &  1.04$\pm$0.12 & \\
                      &$4_{0,4}-3_{0,3}$ & 4-3 \& 5-4  & 37737.318$\pm$0.001&  0.95$\pm$0.25& 5.82$\pm$0.05 & 0.54$\pm$0.08 &  1.64$\pm$0.12 & D\\
                      &$4_{1,3}-3_{1,2}$ & 4-3        & 38672.519$\pm$0.001&  0.53$\pm$0.11& 5.76$\pm$0.08 & 0.80$\pm$0.20 &  0.62$\pm$0.12 & \\
                      &$4_{1,3}-3_{1,2}$ & 3-2 \& 5-4 & 38672.669$\pm$0.001&  0.82$\pm$0.12& 5.85$\pm$0.05 & 0.70$\pm$0.10 &  1.11$\pm$0.12 & D\\
                      &$5_{1,5}-4_{1,4}$ & Main       & 46066.040$\pm$0.002&  1.38$\pm$0.09& 6.01$\pm$0.03 & 0.97$\pm$0.07 &  1.33$\pm$0.13 & \\
                      &$5_{0,5}-4_{0,4}$ & Main       & 47145.610$\pm$0.003&  1.95$\pm$0.14& 5.92$\pm$0.03 & 0.91$\pm$0.08 &  2.02$\pm$0.20 & \\
                      &$5_{1,4}-4_{1,3}$ & Main       & 48333.801$\pm$0.002&  1.19$\pm$0.14& 6.13$\pm$0.06 & 0.89$\pm$0.11 &  1.25$\pm$0.22 & \\
\\
CH$_2$CDCN            &$4_{1,4}-3_{1,3}$ & 4-3        & 36186.447$\pm$0.002&  0.19$\pm$0.07& 5.76$\pm$0.07 & 0.44$\pm$0.27 &  0.41$\pm$0.12 & F\\
                      &$4_{1,4}-3_{1,3}$ & 3-2 \& 5-4 & 36186.600$\pm$0.002&  ...          & ...           & ...           &  ...           & D, I\\
                      &$4_{0,4}-3_{0,3}$ & Main       & 37227.628$\pm$0.002&  0.97$\pm$0.08& 5.87$\pm$0.05 & 0.90$\pm$0.13 &  0.73$\pm$0.09 & \\
                      &$4_{1,3}-3_{1,2}$ & 4-3        & 38369.798$\pm$0.002&  0.31$\pm$0.17& 5.73$\pm$0.12 & 0.70$\pm$0.28 &  0.42$\pm$0.09 & \\
                      &$4_{1,3}-3_{1,2}$ & 3-2 \& 5-4 & 38369.948$\pm$0.002&  0.66$\pm$0.07& 5.82$\pm$0.05 & 1.03$\pm$0.08 &  0.60$\pm$0.09 & D\\
                      &$5_{1,5}-4_{1,4}$ & Main       & 45221.454$\pm$0.002&  0.94$\pm$0.08& 5.90$\pm$0.04 & 1.06$\pm$0.11 &  0.84$\pm$0.11 & \\
                      &$5_{0,5}-4_{0,4}$ & Main       & 46487.192$\pm$0.002&  0.59$\pm$0.09& 5.93$\pm$0.06 & 0.80$\pm$0.16 &  0.70$\pm$0.13 & \\
                      &$5_{1,4}-4_{1,3}$ & Main       & 47949.865$\pm$0.002&  0.16$\pm$0.07& 6.43$\pm$0.03 & 0.24$\pm$0.57 &  0.61$\pm$0.16 & H\\
\\
HCDCHCN               &$4_{1,4}-3_{1,3}$ & 4-3        & 35817.812$\pm$0.021&  0.23$\pm$0.05& 6.09$\pm$0.06 & 0.52$\pm$0.17 &  0.41$\pm$0.09 & F\\
                      &$4_{1,4}-3_{1,3}$ & 3-2 \& 5-4 & 35817.965$\pm$0.021&  0.28$\pm$0.06& 5.96$\pm$0.07 & 0.62$\pm$0.15 &  0.43$\pm$0.09 & D\\
                      &$4_{0,4}-3_{0,3}$ & Main       & 36806.410$\pm$0.041&  0.91$\pm$0.12& 5.83$\pm$0.05 & 0.84$\pm$0.13 &  1.02$\pm$0.15 & \\
                      &$4_{1,3}-3_{1,2}$ & 4-3        & 37882.137$\pm$0.021&  0.22$\pm$0.15& 5.44$\pm$0.14 & 0.60$\pm$0.28 &  0.34$\pm$0.12 & \\
                      &$4_{1,3}-3_{1,2}$ & 3-2 \& 5-4 & 37882.286$\pm$0.021&  ...          & ...           & ...           &  ...           & D, I\\
                      &$5_{1,5}-4_{1,4}$ & Main       & 44762.173$\pm$0.036&  0.56$\pm$0.19& 5.98$\pm$0.12 & 0.72$\pm$0.18 &  0.73$\pm$0.12 & H\\
                      &$5_{0,5}-4_{0,4}$ & Main       & 45967.195$\pm$0.070&  1.07$\pm$0.21& 5.98$\pm$0.06 & 0.61$\pm$0.12 &  1.64$\pm$0.16 & H\\
                      &$5_{1,4}-4_{1,3}$ & Main       & 47341.909$\pm$0.036&  0.27$\pm$0.19& 6.10$\pm$0.32 & 0.82$\pm$0.58 &  0.31$\pm$0.17 & F, H\\
\\
DCHCHCN               &$4_{1,4}-3_{1,3}$ & 4-3        & 34701.375$\pm$0.014&  0.07$\pm$0.03& 5.80$\pm$0.07 & 0.33$\pm$1.37 &  0.21$\pm$0.09 & F\\
                      &$4_{1,4}-3_{1,3}$ & 3-2 \& 5-4 & 34701.527$\pm$0.014&  ...          & ...           & ...           &  ...           & D, I\\
                      &$4_{0,4}-3_{0,3}$ & Main       & 35480.289$\pm$0.019&  ...          & ...           & ...           &  $\le$0.60     & \\
                      &$4_{1,3}-3_{1,2}$ & 4-3        & 36302.640$\pm$0.014&  0.19$\pm$0.05& 6.40$\pm$0.07 & 0.50$\pm$0.13 &  0.35$\pm$0.08 & F\\
                      &$4_{1,3}-3_{1,2}$ & 3-2 \& 5-4 & 36302.789$\pm$0.014&  0.33$\pm$0.06& 6.49$\pm$0.07 & 0.78$\pm$0.16 &  0.40$\pm$0.08 & F, D\\
                      &$5_{1,5}-4_{1,4}$ & Main       & 43371.731$\pm$0.029&  0.59$\pm$0.09& 6.32$\pm$0.08 & 0.98$\pm$0.14 &  0.57$\pm$0.13 & \\
                      &$5_{0,5}-4_{0,4}$ & Main       & 44330.192$\pm$0.037&  1.12$\pm$0.13& 6.01$\pm$0.05 & 0.89$\pm$0.13 &  1.19$\pm$0.20 & \\
                      &$5_{1,4}-4_{1,3}$ & Main       & 45373.023$\pm$0.029&  0.63$\pm$0.18& 6.42$\pm$0.19 & 1.26$\pm$0.44 &  0.47$\pm$0.22 & F\\
\\
CH$_2$CHC$^{15}$N     &$4_{1,4}-3_{1,3}$ &            & 35955.662$\pm$0.003&  0.46$\pm$0.07& 5.62$\pm$0.06 & 0.74$\pm$0.12 &  0.59$\pm$0.09 & \\
                      &$4_{0,4}-3_{0,3}$ &            & 36795.189$\pm$0.003&  ...          & ...           & ...           &  ...           & J\\
                      &$4_{1,3}-3_{1,3}$ &            & 37685.857$\pm$0.003&  0.46$\pm$0.10& 5.82$\pm$0.09 & 0.83$\pm$0.21 &  0.52$\pm$0.13 & \\
                      &$5_{1,5}-4_{1,4}$ &            & 44938.534$\pm$0.005&  0.40$\pm$0.16& 5.64$\pm$0.15 & 0.81$\pm$0.21 &  0.46$\pm$0.20 & F, H\\
                      &$5_{0,5}-4_{0,4}$ &            & 45970.298$\pm$0.007&  0.28$\pm$0.09& 5.87$\pm$0.07 & 0.44$\pm$0.17 &  0.59$\pm$0.17 & \\
                      &$5_{1,4}-4_{1,3}$ &            & 45970.298$\pm$0.007&  ...          & ...           & ...           &  ...           & I\\

\end{longtable}
\tablefoot{
\tablefoottext{a}{Rotational quantum numbers.}\\
\tablefoottext{b}{Total angular momentum including the spin of N. When "Main" is indicated, it refers to the
collapsed hyperfine components with $F_u-F_l$=$J+1$$\rightarrow$$J$,
$J$$\rightarrow$$J-1$, and $J-1$$\rightarrow$$J-2$. In the latter case, the rest frequency corresponds to
that of the strongest component.}\\
\tablefoottext{c}{Rest frequencies used to derive the v$_{LSR}$. If the uncertainty on v$_{LSR}$ is zero, then
the frequencies correspond to those derived assuming v$_{LSR}$=5.83 km\,s$^{-1}$ \citep{Cernicharo2020}.}\\
\tablefoottext{d}{Integrated line intensity in mK\,km\,s$^{-1}$.}\\
\tablefoottext{e}{Velocity of the line with respect to the local standard of rest in km\,s$^{-1}$.}\\
\tablefoottext{f}{Line width at half-intensity using a Gaussian fit in the line profile (in km~s$^{-1}$).}\\
\tablefoottext{g}{Antenna temperature (in mK). Upper limits correspond to 3$\sigma$ values.}\\
\tablefoottext{A}{The line is partially blended with an unknown feature (see Fig. \ref{fig:CH3CH2CCH}). The
line width has been fixed.}\\
\tablefoottext{B}{Data only from the set with frequency-switching throw of 10 MHz.}\\
\tablefoottext{C}{Blend of the hyperfine components of the same rotational transition. If the uncertainty on the
velocity is zero, then the frequencies were
fixed to the predicted ones and the line width was also fixed. The line parameters are correlated.
The value for the total integrated intensity of the line, however, is well derived.}\\
\tablefoottext{D}{Unresolved hyperfine structure. The frequency of the strongest component was taken as reference.
Hence, the derived velocity correspond to an average of the blended hyperfine components.}\\
\tablefoottext{E}{Data only from the set with frequency switching throw of 8 MHz.}\\
\tablefoottext{F}{Line detected at 3$\sigma$ level. The line parameters are uncertain.}\\
\tablefoottext{G}{Blended with a negative feature produced in the folding of the frequency switching data.}\\
\tablefoottext{H}{Blended with another feature. A fit is still possible, but the derived line parameters are uncertain.}\\
\tablefoottext{I}{Blended with an unknown feature.}\\
\tablefoottext{J}{Blended with CH$_3$CN.}\\
}
\end{tiny}
\twocolumn

\end{appendix}
\end{document}